\documentclass[12pt]{iopart}

\newcommand{\aopd}{\hat{a}^\dagger}
\newcommand{\aop}{\hat{a}}

\newcommand{\bopd}{\hat{b}^\dagger}
\newcommand{\bop}{\hat{b}}

\usepackage{graphicx}

\usepackage{iopams}  
\usepackage{amssymb}  
\usepackage{hhline}

\begin{document}

\title{A silicon-organic hybrid platform for quantum microwave-to-optical transduction}

\author{Jeremy D. Witmer$^1$\footnote{These authors contributed equally to this work.}, Timothy P. McKenna$^1\ddagger$, Patricio Arrangoiz-Arriola$^1$, Rapha\"el Van Laer$^1$, E. Alex Wollack$^1$, Francis Lin$^2$, Alex K.-Y. Jen$^2$, Jingdong Luo$^3$ and Amir H. Safavi-Naeini$^1$}

\address{$^1$Ginzton Laboratory, Stanford University, 348 Via Pueblo Mall, Stanford, CA 94305, USA}
\address{$^2$Department of Materials Science and Engineering, University of Washington, Seattle, WA 98195-2120, USA}
\address{$^3$Department of Chemistry, City University of Hong Kong, Hong Kong SAR, P. R. China}

\ead{safavi@stanford.edu}

\begin{abstract}
Low-loss fiber optic links have the potential to connect superconducting quantum processors together over long distances to form large scale quantum networks. A key component of these future networks is a quantum transducer that coherently and bidirectionally converts photons from microwave frequencies to optical frequencies.  We present a platform for electro-optic photon conversion based on silicon-organic hybrid photonics.  Our device combines high quality factor microwave and optical resonators with an electro-optic polymer cladding to perform microwave-to-optical photon conversion from 6.7 GHz to 193 THz (1558 nm).   The device achieves an electro-optic coupling rate of 330 Hz in a milliKelvin dilution refrigerator environment.  We use an optical heterodyne measurement technique to demonstrate the single-sideband nature of the conversion with a selectivity of approximately 10 dB.  We analyze the effects of stray light in our device and suggest ways in which this can be mitigated.  Finally, we present initial results on high-impedance spiral resonators designed to increase the electro-optic coupling.

\end{abstract}


\maketitle

\section{Introduction}

Over the last two decades tremendous advances have been made in the field of superconducting quantum devices \cite{Devoret2013a,Wendin2017}.  As individual quantum processors become more powerful, the potential of connecting these into a large scale quantum network becomes more attractive \cite{Kimble2008a}.  To build such a network, fiber optic links are a natural candidate for transmitting quantum information over long distances due to their low loss ($<$ 0.2 dB/km) and their immunity to thermal noise at room temperature.  However, a key missing component in this network is a quantum transducer capable of converting single photons from microwave frequencies to optical telecommunications frequencies.  Such a converter should support bidirectional operation, operate with near-unity efficiency, have a sufficiently large bandwidth, and add a minimal amount of noise \cite{Lauk2019,Lambert2019}.

In the last few years there have been a number of experimental demonstrations of microwave-to-optical transduction using diverse approaches such as electro-optomechanics \cite{Andrews2014a,Bagci2014,Forsch2019a}, piezo-optomechanics \cite{ Vainsencher2016,Balram2015b,Jiang2019,Shao2019},  direct electro-optic coupling \cite{Rueda2016,Fan2018}, magnons in yttrium iron garnet \cite{Hisatomi2016}, Rydberg atoms \cite{Vogt2019,Han2018}, and rare-earth doped crystals \cite{Fernandez-gonzalvo2019}.  To date, the most compelling results have been achieved using the electro-optomechanical approach, with conversion efficiencies as high as 47\% \cite{Higginbotham2018}.  However, the direct electro-optic approach also has several advantages: it can be fabricated on a planar substrate, has the potential for large conversion bandwidth ($>$ MHz), and can easily be made voltage tunable.  So far experimental approaches for direct electro-optic conversion have used either lithium niobate (LiNbO$_3$) \cite{Rueda2016,Witmer2017a} or aluminum nitride (AlN) \cite{Fan2018} as the  electro-optic material, but it is not clear whether these are optimal choices moving forward.

One class of electro-optic devices that has achieved impressive results for classical modulators are silicon-organic hybrid (SOH) devices \cite{Leuthold2009,Leuthold2013,Koos2016}, which consist of silicon waveguides clad in an organic electro-optic (EO) polymer \cite{Dalton2010}.  The waveguides are often designed so that light is confined to a narrow polymer-filled slot, giving high overlap between the optical mode and the EO polymer.  If the silicon device layer is made  conductive by doping, then the modulating voltage can be dropped entirely across the narrow slot, resulting in large electric fields and extremely high effective electro-optic coefficients.  For example, the EO polymer used in \cite{Kieninger2018} was shown to have a material Pockels coefficient of 390 pm/V in the modulator device, more than 10 times larger than the Pockels coefficient of lithium niobate ($\approx$ 31 pm/V) \cite{Weis1985}.  This same device achieved a record-setting $V_\pi L = 0.32$ V$\cdot$mm, almost 90 times smaller than the recent work by Wang et al. in thin-film LiNbO$_3$.  (However, the SOH device also had a large optical propagation loss of 3.9 dB/mm, about 200 times larger than the LiNbO$_3$ device.)  In addition to having extremely high EO coefficients, SOH devices have been shown to support high speed modulation ($>$100 GHz) \cite{Korn2013,Alloatti2014,Wolf2018}, have adequate thermal stability (reliable operation above 100 $^{\circ}$C) \cite{Miura2017}, and maintain good performance at cryogenic temperatures (7 K) \cite{Park2016}.

In this work we experimentally explore the suitability of silicon-organic hybrid devices for quantum microwave-to-optical transduction.  We design and demonstrate a sideband-resolved converter and characterize it in a dilution refrigerator, and suggest future steps to improve the transduction efficiency.

\section{Device overview and theory of operation}

\begin{figure} [htbp]
\centering
\includegraphics[width=5 in]{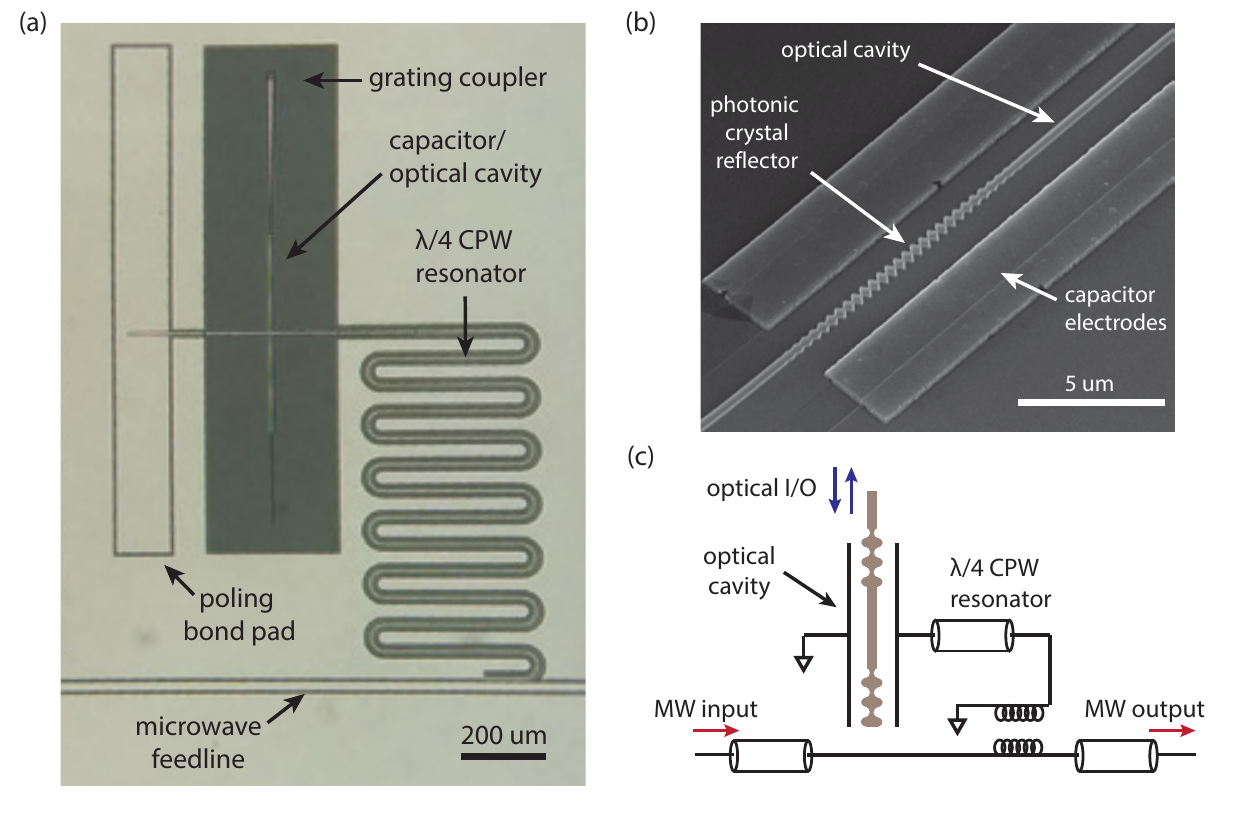}
\caption{(a) An optical micrograph of the electro-optic converter device fabricated on an SOI chip.   (b) A scanning electron microscope (SEM) image showing a zoomed-in view of the optical waveguide between two capacitor electrodes.  An optical cavity is formed by adding photonic crystal reflectors to either end of the waveguide. The images in (a) and (b) were taken before the EO polymer was applied to the chip.  (c) A schematic representation of the electro-optic converter device. }
\label{fig:overview}
\end{figure}

To achieve electro-optic conversion from microwave photons to optical photons, our device combines high quality (Q) factor optical and microwave resonators with an electro-optic polymer cladding.  An overview of the device can be seen in Figure \ref{fig:overview}(a).  The microwave resonator consists of a $\lambda/4$ coplanar waveguide (CPW)  terminated by a capacitor.  The capacitor electrodes induce an electric field which extends across a silicon optical waveguide (Figure \ref{fig:overview}(b)) and changes the refractive index of the EO polymer cladding via the linear electro-optic (Pockels) effect.  Since the optical mode has a large evanescent overlap with the EO polymer, this in turn affects the optical resonance frequency. Figure \ref{fig:overview}(c) shows a schematic representation of the converter device.

The theoretical operation of an electro-optic microwave-to-optical transducer has been laid out in detail in \cite{Tsang2010} and \cite{Tsang2011}.  The Hamiltonian describing the device is 
\begin{equation}
\hat{H} = \hbar \omega_\textrm{\scriptsize opt}\aopd\aop +\hbar \omega_\textrm{\scriptsize MW}\bopd\bop - \hbar g_0 \left(\bop + \bopd\right)\aopd\aop
\end{equation}
where $\omega_\textrm{\scriptsize opt}$ ($\omega_\textrm{\scriptsize MW}$) is the optical (microwave) resonator frequency, $\aop$ ($\bop$) is the annihilation operator for optical (microwave) cavity photons, and $g_0$ is the electro-optic coupling rate.  This Hamiltonian has the same form as the Hamiltonian for optomechanics \cite{Aspelmeyer2014}, and thus supports an itinerant conversion scheme first proposed for optomechanical devices in \cite{Safavi-Naeini2011a} and considered in more detail in \cite{Wang2012}.  A red-detuned optical pump field is applied, the Hamiltonian can be linearized and the rotating-wave approximation can be applied, leading to a beamsplitter-type interaction Hamiltonian
\begin{equation}
    \hat{H}_\textrm{\scriptsize int} = -\hbar g_0 \sqrt{n_\textrm{\scriptsize cav}}\left(\aopd\bop + \aop \bopd\right)
\end{equation}
where $n_\textrm{\scriptsize cav}$ is the mean intracavity optical photon number.  When the pump is red-detuned by exactly the microwave frequency, the photon number conversion efficiency from microwaves to optics is given by \cite{Tsang2011}
\begin{equation}
    \eta =  \frac{\kappa_e \gamma_e}{\kappa_\textrm{\scriptsize tot} \gamma_\textrm{\scriptsize tot}} \frac{4C}{(1+C)^2}
\end{equation}
where $\kappa_e$ ($\gamma_e$) is the extrinsic optical (microwave) loss rate, $\kappa_\textrm{\scriptsize tot}$ ($\gamma_\textrm{\scriptsize tot}$) is the total optical (microwave) loss rate, and $C = \frac{4g_0^2n_\textrm{\scriptsize cav}}{\kappa_\textrm{\scriptsize tot}\gamma_\textrm{\scriptsize tot}}$ is the cooperativity.  Achieving high conversion efficiency therefore requires over-coupled resonators ($\frac{\kappa_e}{ \kappa_\textrm{\scriptsize tot}}, \frac{\gamma_e}{ \gamma_\textrm{\scriptsize tot}} \approx 1$) and near-unity cooperativity ($C \approx 1$).  To increase the electro-optic cooperativity towards unity, it is necessary to reduce both the optical and microwave loss rates, as well as to maximize the electro-optic interaction strength $\sqrt{n_\textrm{\scriptsize cav}}g_0$.

To better understand the electro-optic coupling rate $g_0$, we can write it as
\begin{eqnarray}
    g_0 &= g_V V_\textrm{\scriptsize zpf} \\
    &= g_V ~\omega_\textrm{\scriptsize MW} \sqrt{\frac{\hbar Z}{2}}
    \label{eqn:g0}
\end{eqnarray}
where $g_V/2\pi$ is the electro-optic tuning rate of the optical resonator measured in units of hertz per volt,  $V_\textrm{\scriptsize zpf}$ is the zero-point voltage fluctuation on the capacitor of the microwave LC resonator, and $Z$ is the microwave resonator impedance.  From first-order perturbation theory \cite{Joannopoulos2008}, the electro-optic tuning rate $g_V$ is given by an integral of the optical field over the EO polymer region
\begin{equation}
g_V = \frac{\Delta\omega_\textrm{\scriptsize opt}}{V_\textrm{\scriptsize applied}} = -\frac{1}{V_\textrm{\scriptsize applied}}\frac{\omega_\textrm{\scriptsize opt}}{2} ~\frac{\sum_{ij} \int_\textrm{\scriptsize polymer} E^*_{0i} \Delta\varepsilon_{ij} E_{0j} ~d^3r}{\sum_{ij} \int E^*_{0i} \varepsilon_{ij} E_{0j} ~d^3r},
\end{equation}
where $E_0 = (E_{0x},E_{0y},E_{0z})$ is the electric field of the unperturbed optical mode, and $\Delta\varepsilon$ is the permittivity change due to the electro-optic effect in the polymer. The primary component of the polymer electro-optic tensor is $r_{33}$, where the `3' direction here is in the plane of the chip, perpendicular to the waveguide. If we include only this component of the electro-optic tensor then we can write $\Delta n = \Delta\varepsilon / 2 n =  -\frac{1}{2} n_\textrm{\scriptsize polymer}^3 r_{33} E_\textrm{\scriptsize RF,3}$, with $E_\textrm{\scriptsize RF,3}$ the radio frequency (RF) electric field from the electrodes.  If we further make an approximation that $\Delta\varepsilon$ is constant across the relevant area of the polymer and that the optical mode is TE polarized, then we can express the electro-optic tuning compactly as
\begin{equation}
    g_V \approx \frac{1}{2}\omega_\textrm{\scriptsize opt} n_\textrm{\scriptsize polymer}^2 r_\textrm{\scriptsize 33} \left(\frac{E_\textrm{\scriptsize RF,3}}{V_\textrm{\scriptsize applied}}\right) \left( \frac{U_\textrm{\scriptsize polymer}}{U_\textrm{\scriptsize total}}\right)
\end{equation}
where $U_\textrm{\scriptsize polymer} / U_\textrm{\scriptsize total}$ is the fraction of the optical mode energy in the polymer.

\section{Fabrication}

\subsection{Fabrication overview}
One of the key advantages of the silicon-organic hybrid  platform is that it can be fabricated almost entirely using standard silicon photonics processing.  The only non-standard steps are a final spin coating of the dies with EO polymer and subsequent poling.

\begin{figure} [htbp]
\centering
\includegraphics[width=5 in]{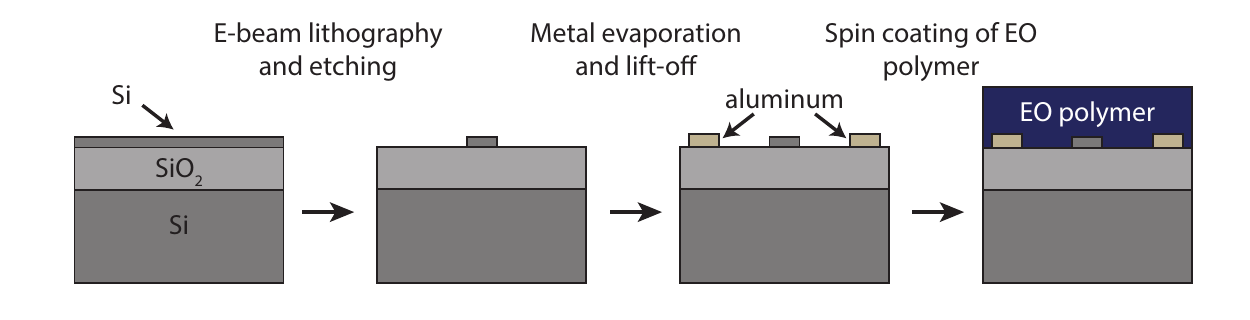}
\caption{A simplified illustration of the nanofabrication process for the electro-optic converter device.  }
\label{fig:fab}
\end{figure}

A simplified fabrication flow is shown in Figure \ref{fig:fab}.  The fabrication begins with an unpatterned silicon-on-insulator (SOI) die with a 220 nm thick Si device layer and a 3 $\mu$m buried oxide.  First, the silicon waveguides are patterned using electron beam lithography (100 keV JEOL system, CSAR resist) and a Cl$_2$/HBr reactive ion etch.  Next, a photolithography step (SPR-3612 resist) and second reactive ion etch is used to remove the remainder of the device layer silicon across the die, exposing the buried oxide.  The first metallization step uses photolithography (AZ-5412 resist) to define the ground plane and larger features, followed by the evaporation and lift-off of 100 nm of aluminum. The second metallization step uses election beam lithography (MMA/PMMA resist bilayer) for precise definition of the electrodes, followed by evaporation and lift-off of 200 nm of aluminum.  After the metallization, the dies are ready for the EO polymer to be applied.  We use SEO100C from Soluxra as our electro-optic polymer because of its large electro-optic coefficient (approximately 105 pm/V) \cite{Huang2012}. The polymer is first dissolved in cyclopentanone  and then spin coated onto the dies and baked dry, resulting in a polymer cladding thickness of 500-1000 nm.  

\subsection{Poling}
\label{sec:poling}
\begin{figure} [htbp]
\centering
\includegraphics[width=5 in]{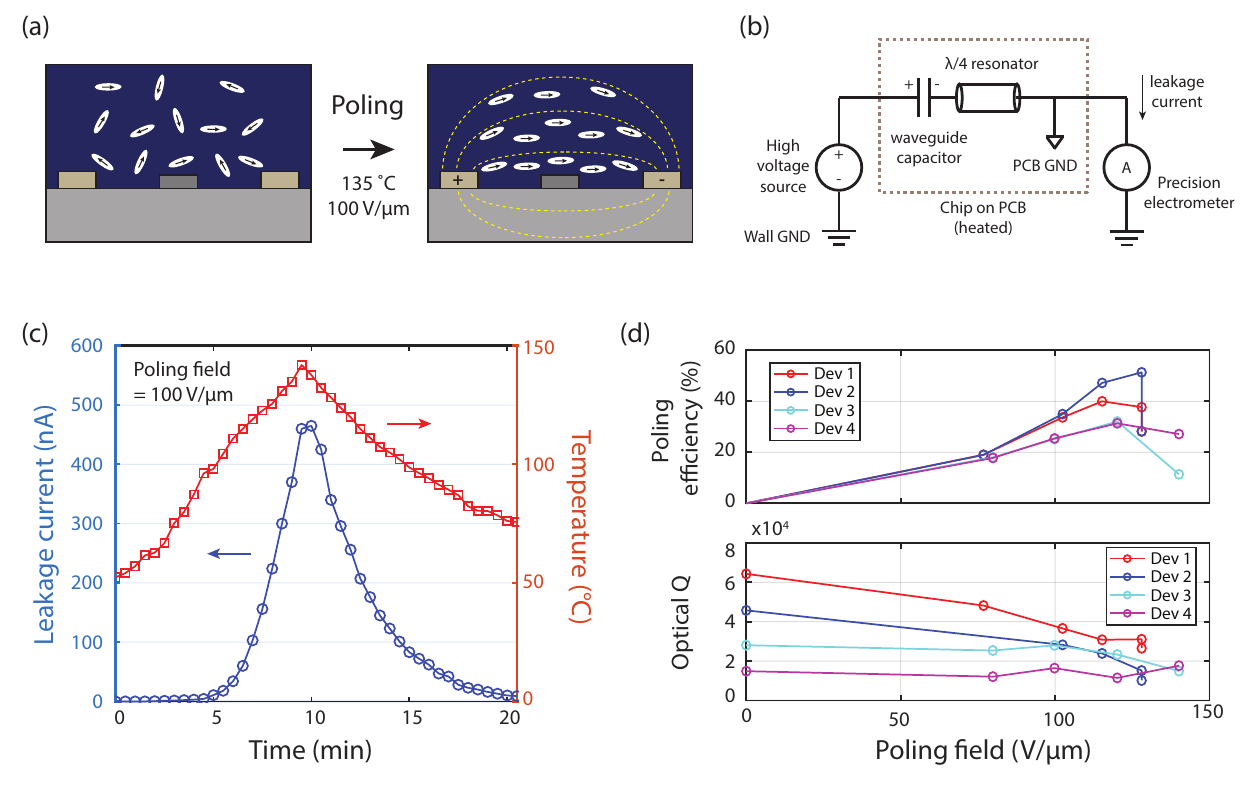}
\caption{(a) The EO polymer SEO100C consists of electro-optic chromophore molecules (white ellipses) embedded in a host matrix (dark blue).  After spin coating the chromophore molecules are randomly oriented.  The poling process orients the chromophores along the electric field between the electrodes. The chromophore molecules are not shown to scale, the actual size is roughly 1--2 nm.  (b) Circuit diagram showing the in-device poling procedure for the EO polymer. (c) shows the approximate die temperature and measured leakage current during the poling process.  As the die approaches the glass transition temperature of the polymer (135 $^\circ C$), a large spike in the leakage current is observed. The hotplate is turned off after the glass transition temperature is reached and the die is allowed to cool.  (d) shows the effects of varying the poling field on the poling efficiency and total optical Q factor.  Four different devices were repeatedly poled under gradually increasing bias fields.  The devices were cooled to room temperature and measured between each trial.  The optical Q's at zero poling field were measured after spin-coating the die with EO polymer, but before any poling. The lines connecting points in (c) and (d) are guides to the eye.}
\label{fig:poling}
\end{figure}

When the EO polymer is first applied to the dies, the polar chromophore molecules are randomly oriented (Figure \ref{fig:poling}(a)).  In order to produce a net EO effect in the device it is necessary to align the chromophore molecules along the direction of the electric field between the electrodes.  To do this we mount the die on a printed circuit board (PCB) and apply a large bias voltage directly across the capacitor via the poling bond pad shown in Figure \ref{fig:overview}(a). A circuit diagram illustrating the electrical connections during the poling process is shown in Figure \ref{fig:poling}(b). Typical bias voltages are around 300 V, resulting in an electric field of approximately 100 V/$\mu$m in the device.  While the bias voltage is being applied, we use a hotplate to gradually increase the die temperature until it reaches the EO polymer glass transition temperature of approximately 135 $^{\circ}$C.  During this process we use a precision electrometer (Keithley 617) to monitor the leakage current through the polymer.  As shown in Figure \ref{fig:poling}(c), the leakage current increases rapidly as the die approaches the glass transition temperature. Once the target temperature is reached, the die is allowed to cool with the bias voltage still applied. When poled, the EO polymer is birefringent with extraordinary refractive index $n_e = 1.70$ along the poling direction and ordinary refractive index $n_o = 1.65$ perpendicular to the poling direction. 

After the poling is finished, the device EO tuning can be measured and compared to simulations to extract the effective in-device EO coefficient of the polymer film.  In practice, the measured EO coefficient of the polymer in a modulator device is always lower than the maximum EO coefficient obtained with uniform thin films \cite{Heni2017}.  This is thought to be due to a variety of effects such as non-uniformity of the poling field, geometric filling effects, and interaction between chromophore molecules and waveguide surfaces \cite{Heni2017}.  The ratio of the in-device material EO coefficient to the thin-film material EO coefficient is known as the poling efficiency \cite{Chen2008a}.  In SOH modulator devices there is often a trade-off between a larger poling field (which can increase poling efficiency) and optical loss \cite{Chen2008a}, and we observe this in our devices as well.  Figure \ref{fig:poling}(d) shows the measured poling efficiency and optical Q factor for four different devices under different applied poling fields.  The devices were repoled multiple times with increasing bias voltage (with the same polymer cladding), and the devices were cooled to room temperature and measured between each poling.  As expected, a larger poling field causes the optical Q to gradually decrease as poling induced losses grow.  On the other hand, a larger poling field is also observed to improve the poling efficiency up to a maximum of roughly 30--50\% at 120 V/$\mu$m.  The trade-off illustrated here led us to adopt a poling field of 100 V/$\mu$m for the majority of our devices.

\section{Optical design and characterization}

\subsection{Slotted vs. unslotted waveguides}

\begin{figure} [htbp]
\centering
\includegraphics[width=\textwidth]{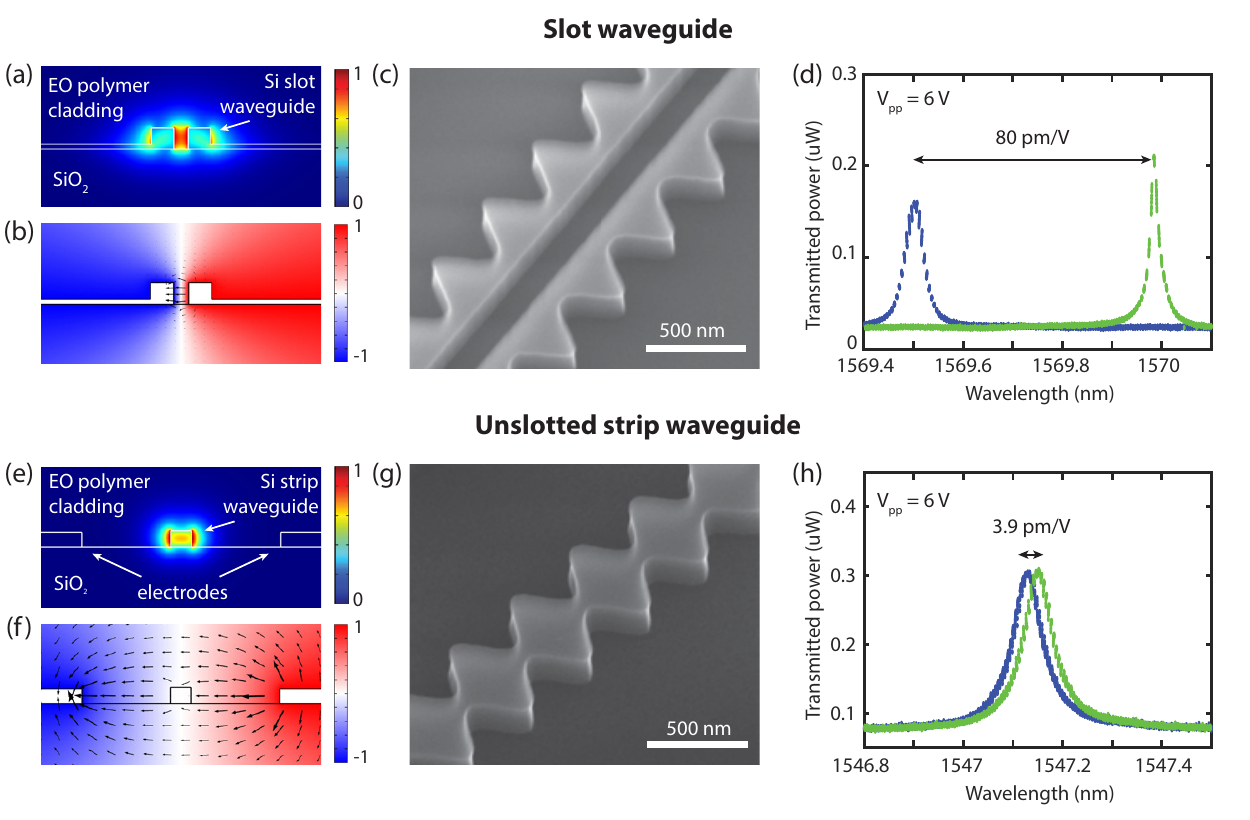}
\caption{(a) and (e) show  COMSOL simulations of the optical waveguide TE mode for a slotted and unslotted waveguide geometry, respectively.  The color shows the magnitude of the electric field.  (b) and (f) show COMSOL simulations of the RF electric field for the two waveguide geometries.  For a strip-loaded slot waveguide the electric field is confined almost entirely to the slot region, while for the unslotted waveguide with external electrodes the electric field is much more extended.  (c) and (g) show SEM images of a slotted and unslotted waveguide.  The waveguide widths are periodically modulated to create a photonic crystal reflector.  The device shown in (g) has a fully etched silicon device layer (it does not have the conductive silicon slab pictured in the simulations).  These images were taken before EO polymer was applied to the chip.  (d) and (h) show plots comparing the measured DC tuning of the optical resonance for the two devices.  For the slot waveguide the modulation voltage drops across the 180 nm slot in the center of the waveguide, while for the unslotted waveguide the voltage drops across two external electrodes (not shown) with a separation of 3 $\mu$m.  The peak-to-peak voltage in both cases is 6 V.  The stroboscopic technique used to measure the EO tuning is described in Section \ref{sec:optical_characterization}.}
\label{fig:slotted}
\end{figure}

Generally speaking, the largest device EO coefficients in SOH modulators are achieved using a so-called strip-loaded slot waveguide geometry \cite{Leuthold2013}, which is illustrated in Figure \ref{fig:slotted}(a) and (b).  In this configuration, the optical field is confined primarily in the polymer-filled slot between two silicon strips and the modulating voltage is dropped directly across the slot.
However, for this geometry to work as described, the silicon slab on either side of the waveguide must be made sufficiently conductive through doping so that it can transmit the voltage on the electrodes. In comparison, an unslotted strip waveguide with adjacent electrodes (Figure \ref{fig:slotted}(e) and (f)) will typically have a device EO coefficient that is smaller by a factor of 10 or more owing to the reduced mode overlap with the EO polymer and the much larger spacing between the electrodes, but it also does not require silicon doping.

To initially evaluate the EO performance of the slotted waveguide geometry, we fabricated fully-etched slotted photonic crystal cavities, pictured in Figure \ref{fig:slotted}(c).   To convert from strip feed waveguides to a slotted waveguide we used mode converters similar to the one in \cite{Zhang2015}.  For these test devices the slotted waveguides did not have any strong doping or partially etched silicon slab; instead, we deposited metal leads that directly contacted each side of the slotted waveguide at the far tip of the mode converters.  These optical cavities were measured to have total Q factors ranging from about 30,000 to 60,000 after poling, and a large EO tuning coefficient of $\approx$ 80 pm/V (Figure \ref{fig:slotted}(d); see Section \ref{sec:optical_characterization} for the measurement procedure).  This is much higher than the tuning of 2--6 pm/V that we observed for our unslotted strip waveguides (Figure \ref{fig:slotted}(e)).  However, because the slotted waveguide tuning relied on the conductivity of the lightly p-doped silicon device layer (resistivity of 10--15 $\Omega\cdot$cm), the modulation had a 3 dB roll-off around 20 kHz making it unsuitable for microwave signals.

\begin{figure} [htbp]
\centering
\includegraphics[width=\textwidth]{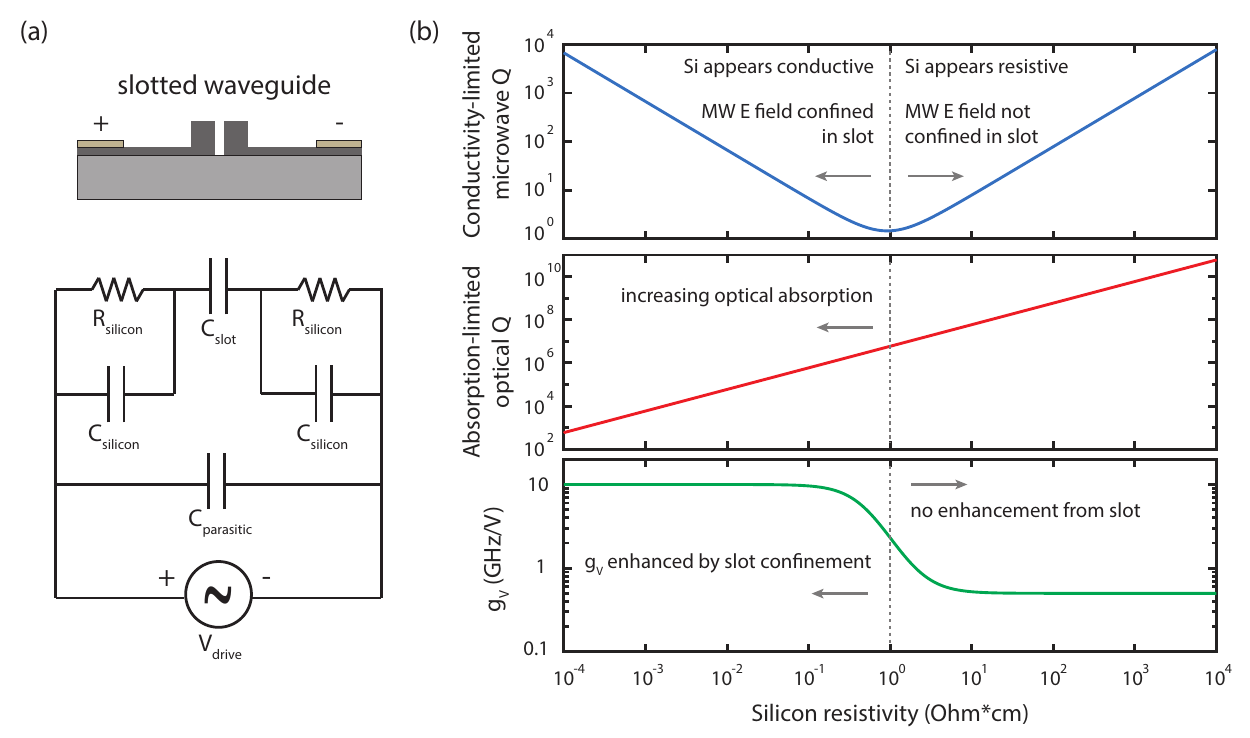}
\caption{(a) Equivalent circuit of the strip-loaded slot waveguide geometry.  (b) Plots showing how the simulated microwave Q, optical Q and resonator tuning $g_\textrm{\scriptsize V}$ change with the silicon device layer resistivity.  The microwave Q is estimated by taking the ratio of the imaginary and real parts of the impedance for the circuit in (a) at a frequency of 5 GHz.  The optical Q is estimated based on optical absorption data for n-doped silicon in  \cite{Degallaix2013}. $g_\textrm{\scriptsize V}$ is assumed to be directly proportional to the voltage across $C_\textrm{\scriptsize slot}$ in (a). The device simulated here had a silicon slab height of 50 nm, a waveguide height of 220 nm, a slot width of 150 nm, and an electrode separation of 2.15 $\mu$m.  Note that these plots are meant to provide intuition and illustrate design trade-offs, not necessarily to capture all the effects present in a real device.}
\label{fig:doping}
\end{figure}

Figure \ref{fig:doping} illustrates the key trade-offs for a strip-loaded slot waveguide modulator operating as part of a quantum transducer.  In order to take advantage of the large device EO coefficient afforded by the slot geometry, the silicon slab must be doped to reduce the resistivity (for the representative device geometry simulated here, the requirement was resistivity $\rho \lesssim 0.2 ~\Omega\cdot$cm).  However, if the modulator is integrated as part of a microwave resonator, then the microwave Q also depends critically on the silicon slab resistivity.  High microwave Q's ($> 1000$) are only possible for very large resistivities ($\gtrsim 1000~\Omega\cdot$cm) or very small resistivities ($\lesssim 10^{-3}~\Omega\cdot$cm). Moreover, if the resistivity is too small then optical absorption in the silicon will start to limit the optical Q.  Overall, it is difficult to find a slot waveguide geometry and doping level which simultaneously give high microwave and optical Q's and also keep the microwave electric field confined to the slot.  This is not a major issue for classical EO modulators, where microwave and optical loss are somewhat less critical (a detailed discussion of the relative advantages of slotted and unslotted waveguides for classical modulators can be found in \cite{Leuthold2013}).  However, since quantum transducers require high cooperativity and sideband-resolved operation, we decided to use an unslotted waveguide geometry for the EO converter device presented in this paper.

\subsection{Optical cavity design}

\begin{figure} [htbp]
\centering
\includegraphics[width=5 in]{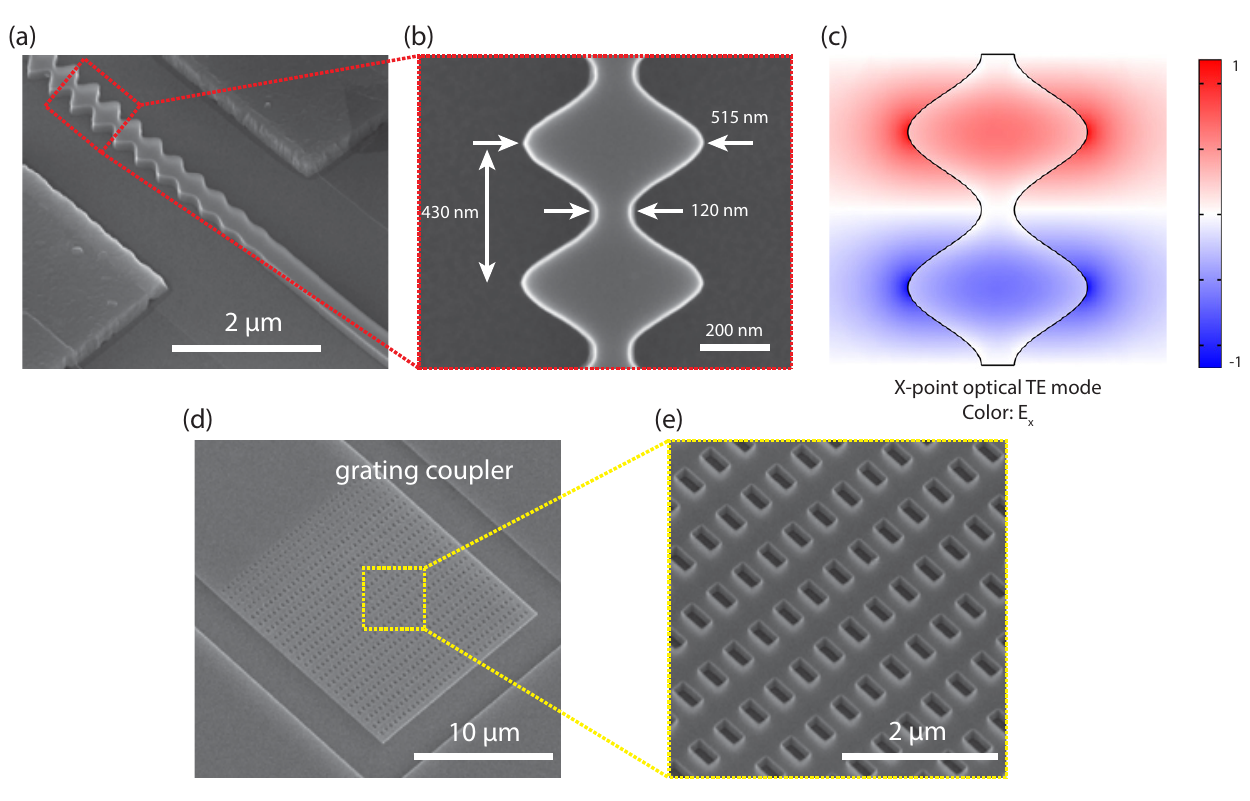}
\caption{(a) An SEM image showing the optical waveguide with photonic crystal reflector between two electrodes. (b) A zoomed-in image of two unit cells in the ``fishbone'' reflector.  (c) A COMSOL simulation of the optical TE mode in the structure, at the X-point of the dielectric band.  The color shows the x component of the electric field. (d) An SEM image of the grating coupler. The grating is formed by a regular array of rectangular holes etched into the silicon slab. (e) A zoomed-in view of the etched holes in the grating coupler. These SEM images were taken before EO polymer was applied to the chip.}
\label{fig:optical_design}
\end{figure}

To confine light in three dimensions on the chip we use a silicon waveguide terminated with a pair of photonic crystal mirrors (Figure \ref{fig:optical_design}(a)).  Rather than forming the photonic crystal by etching holes into the center of the silicon waveguide \cite{Witmer2016b,Witmer2017a,Deotare2009}, we instead open up a band-gap by sinusoidally modulating the waveguide width, similar to \cite{Goban2015}.  This type of modulation, which we call a ``fishbone'' geometry, has two main advantages.  First, it removes the need to etch very narrow holes, which can be difficult to fabricate.  Second, because the geometry perturbation is happening at the edges of the waveguide where the optical field is much smaller  than in the center of the waveguide, we can achieve smaller effective index perturbations than would be possible with etched holes.  This allows us to adjust the strength of the index perturbation very gradually and therefore create an adiabatic transition from the unperturbed waveguide to the photonic crystal region. 

The period and modulation strength of the photonic crystal unit cell are designed to give a large bandgap of approximately 60 nm at a center wavelength around 1550 nm.  The number of mirror periods in the reflectors is chosen to achieve a desired extrinsic coupling Q, typically around 50,000.  Typical devices use 5 to 10 full strength mirror periods with an additional 10 periods on either side in which the modulation is linearly ramped down to make the transition more adiabatic. The unperturbed waveguide is 290 nm wide and 220 nm tall, which results in approximately 35\% of the optical mode energy being confined in the EO polymer.   Typical dimensions for the reflector unit cells are shown in Figure \ref{fig:optical_design}(b), and the TE optical mode profile for the dielectric band at the X-point is shown in Figure \ref{fig:optical_design}(c).  The length of the unperturbed waveguide between the reflectors is 420 $\mu$m.  

There is a trade-off in determining how the electrodes should be positioned around the waveguide \cite{Witmer2016b}.  Decreasing the gap between the electrodes will result in a larger electric field per volt (and therefore a larger $g_V$), but will eventually also increase the optical loss due to absorption in the metal.  We chose a gap of 2.7 $\mu$m to maximize $g_V$ while keeping the simulated absorption loss below $\approx$ 0.1 dB/cm.

\subsection{Optical coupling}

In order to couple light on and off chip, we use meta-surface grating couplers based on the design in \cite{Benedikovic2014}, shown in Figure \ref{fig:optical_design}(d).  The grating consists of a regular array of rectangular holes (Figure \ref{fig:optical_design}(e)) which create a periodic refractive index perturbation to scatter the light from free space into the waveguide.  The first four rows of holes are slightly smaller than the rest, which improves index matching between the grating coupler and the waveguide, reducing back reflections. These grating couplers have a FWHM transmission band of approximately 40 nm, with typical peak efficiencies around 30\%. We send light to the grating couplers using angle polished fibers, which we can also glue to the chip to achieve reliable cryogenic packaging using the method described in \cite{McKenna2019}.  After the grating coupler, the waveguide is tapered from an initial width of 15 $\mu$m to a final width of 290 nm over a distance of about 300 $\mu$m.

\subsection{Optical characterization}
\label{sec:optical_characterization}

\begin{figure} [htbp]
\centering
\includegraphics[width=5 in]{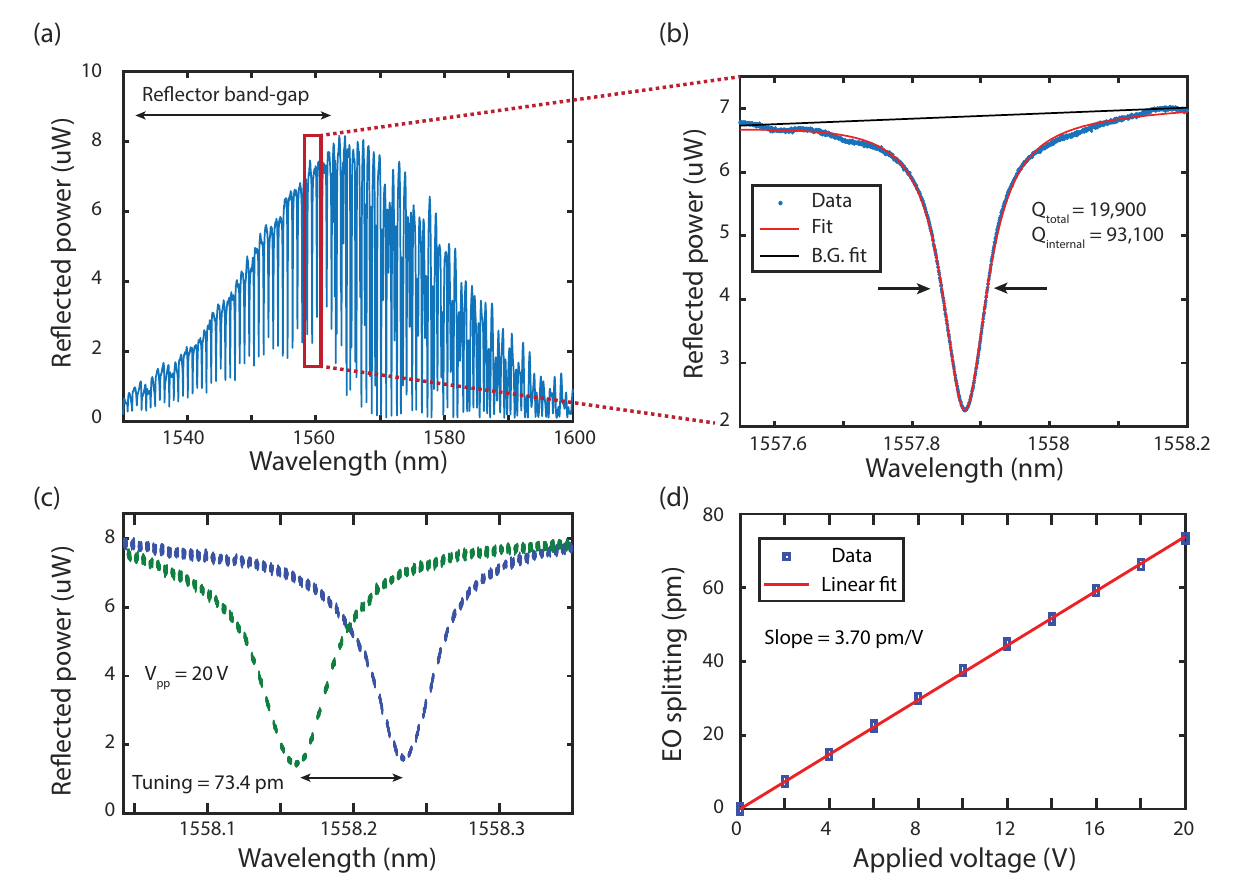}
\caption{(a) The reflected power spectrum from the optical cavity, measured at a temperature of approximately 7 mK.  The overall Gaussian shape is due to the transmission spectrum of the grating coupler with glued fibers.  The left half of the spectrum (below 1565 nm) lies within the band-gap of the photonic crystal reflectors, and so the reflection spectrum consists of a set of closely spaced resonances.  (b) shows a zoomed-in spectrum of the optical resonance used for the EO conversion experiment.  The resonance dip is fit using a Fano-Lorentzian lineshape (red) with a linear background (black).  (c) shows the same resonance as (b), but measured at room temperature.  The electrodes are modulated with a 20 V$_\textrm{\scriptsize pp}$ square wave at 200 Hz, and the laser wavelength is scanned at 1 nm/s so the resonance appears at two different locations in the spectrum.  The two dips are fit independently to extract the splitting.  (d) shows the extracted splitting values for the resonance in (c) as a function of the applied peak-to-peak voltage.  The extracted electro-optic tuning rate is 3.70 pm/V.}
\label{fig:optics}
\end{figure}

We characterize the optical performance of the device by scanning a tunable laser (Santec TSL-550) and monitoring the reflected power.  Figure \ref{fig:optics}(a) shows a typical reflection spectrum, taken with the device in the dilution refrigerator.  The overall Gaussian shape is determined by the grating coupler response.  The left hand side of the spectrum (below 1565 nm) lies within the band-gap of the photonic crystal reflectors, resulting in a series of Fabry-Perot resonances with a free-spectral range of 2.3 nm.  The right hand side of the spectrum (above 1565 nm) is outside of the band-gap, and shows high frequency ripples due to spurious reflections within the device. Figure \ref{fig:optics}(b) shows a zoomed-in view of one of the resonances.  Fitting this to a Fano-Lorentzian we extract an intrinsic Q of 93,100 and a total Q of 19,900.  The extrinsic Q is 25,300, so the device is somewhat overcoupled.

We use a stroboscopic technique to quantify the electro-optic performance of the device at low frequencies.  We directly apply a voltage signal from an arbitrary waveform generator (Rigol DG4102) across the waveguide capacitor by wirebonding to the poling bond pad (shown in Figure \ref{fig:overview}(a)).  (For subsequent RF measurements in the dilution refrigerator, this bond pad is shorted to the chip ground with wirebonds.)  Scanning the laser slowly across the resonance while applying a square wave voltage to the device, we see the Lorentzian dip appear at two different positions due to the electro-optic shift.  In the case pictured in Figure \ref{fig:optics}(c) the laser scan rate was 1 nm/s and the square wave frequency was 200 Hz.  We also stepped the modulation frequency up to 10 MHz (the limit of the arbitrary waveform generator) and observed no significant decrease in the tuning.  An advantage of measuring electro-optic tuning in this way compared to a DC tuning measurement is that it is self-referenced, removing any effects of hysteresis in the device or wavelength jitter from scan to scan.  Figure \ref{fig:optics}(d) shows the EO tuning as a function of applied peak-to-peak voltage, leading to an extracted tuning rate of 3.7 pm/V for this device.  We characterized the EO tuning for 9 devices like this on 4 different chips and found that the tuning rates were consistently between 2 and 6 pm/V, depending on the electrode spacing and poling conditions (previously discussed in Section \ref{sec:poling}).

In the process of packaging the chips for cryogenic measurement, we noticed that the EO coefficient decreased dramatically after fibers were glued onto the chips, but before the chips were cooled down.  For example, for the device described in Figure \ref{fig:optics}, the EO tuning decreased from 3.7 pm/V to 1.1 pm/V after gluing.  During the gluing, the chips are subjected to an intense UV cure for up to 8 minutes \cite{McKenna2019}, during which time the chips can become hot.  We believe the elevated temperature during the UV curing process caused the EO polymer film to become partially depoled or photochemically bleached, reducing the EO coefficient.  Since the EO conversion efficiency scales like $g_0^2$, we estimate that this reduced our overall conversion efficiency by approximately a factor of 11.  We believe this issue can be addressed in the future by selecting a different optical adhesive with better chemical compatibility with EO polymers.

\section{Microwave design and characterization}
\label{sec:microwave}

\subsection{Microwave design}
The efficiency of an electro-optic transducer can be increased by recirculating the microwave photons in a resonator \cite{Tsang2011}.  In our device, we use a $\lambda/4$ CPW resonator, as shown in Figure \ref{fig:overview}(a), with one end shorted to the chip ground plane and the other terminated by a capacitor which spans the optical cavity. The silicon device layer has been removed in the metallized regions, so the  CPW lies directly on top of the silicon oxide layer.  The CPW resonator has a 5 $\mu$m center conductor width with a 13 $\mu$m gap, which leads to an impedance of approximately 100 $\Omega$.   The CPW resonators are inductively coupled to the 50 $\Omega$ microwave feedline, with coupling rates designed to be close to the expected intrinsic loss rates. 

The CPW has a length of 5200 $\mu$m and a simulated inductance (capacitance) per unit length of 620 nH/m (63 pF/m).  The capacitor at the open end of the CPW consists of two 3 $\mu$m wide electrodes with a gap of 2.7 $\mu$m and a length of 450 $\mu$m, and it has a simulated capacitance of 21 fF.

\subsection{Microwave characterization}

\begin{figure} [htbp]
\centering
\includegraphics[width=5 in]{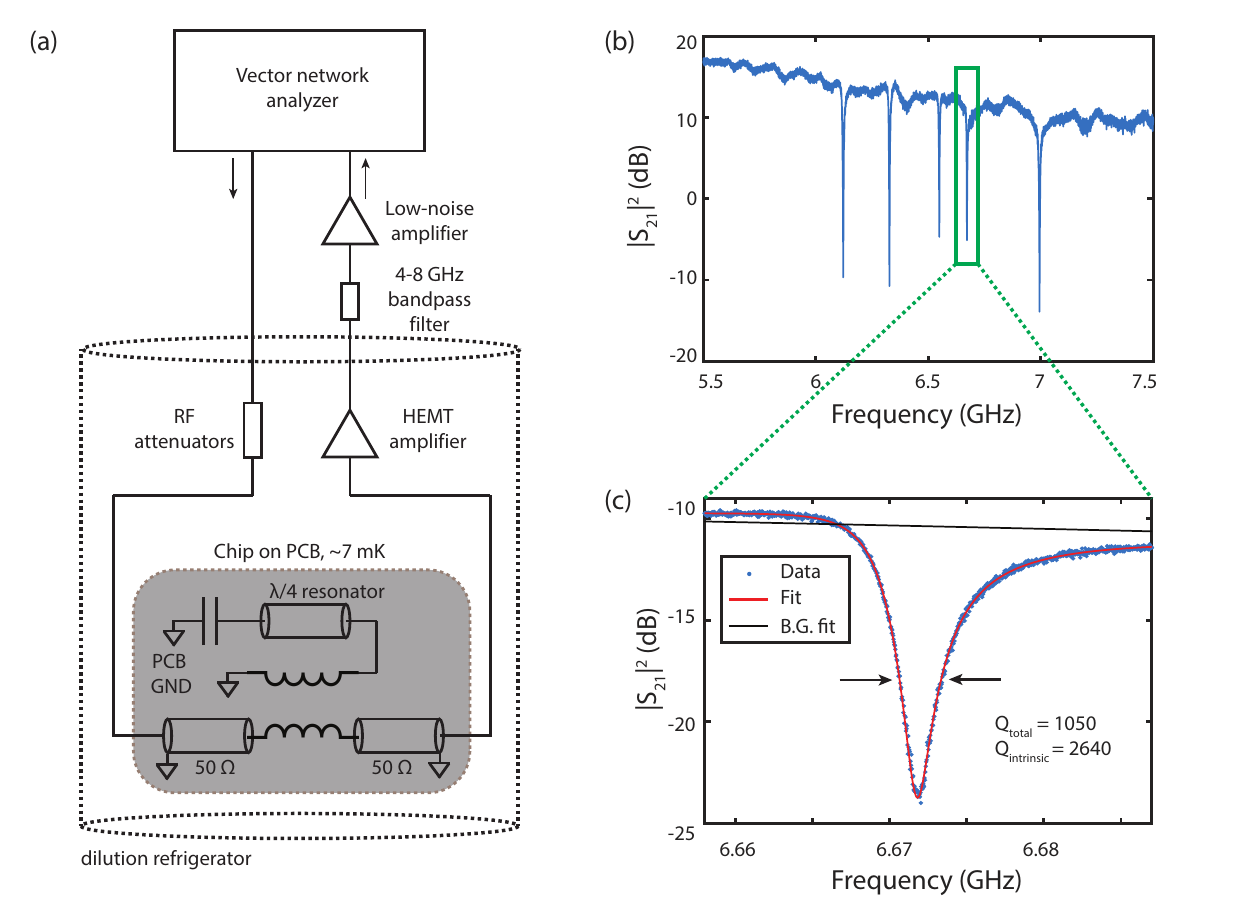}
\caption{(a) A schematic of the measurement setup used to characterize the microwave resonators in the dilution refrigerator.  (b) The microwave transmission spectrum of the EO converter chip. The five resonances between 6 and 7 GHz correspond to five different CPW devices on the chip, side-coupled to the same feedline.   (c) A zoomed-in view of the microwave mode used for EO conversion, fit with a Fano-Lorentzian lineshape (red) and linear background (black). This spectrum was taken using an on-chip RF excitation power of $-140$ dBm. The offset of the y-axis in (c) compared to (b) is due to the presence of an extra 20 dB attenuator outside the fridge for this particular measurement. }
\label{fig:MW_spectrum}
\end{figure}

To characterize the microwave resonators, we wirebond the chips into a PCB and cool them to a temperature of approximately 7 mK in a Bluefors dilution refrigerator and measure the $S_{21}$ scattering parameter using a vector network analyzer (VNA, Rohde \& Schwarz ZNB20).  The measurement setup is illustrated in Figure \ref{fig:MW_spectrum}(a).  We use 50 dB of attenuation on the input lines to reduce thermal noise in the microwave mode, and amplify the output signal using a HEMT amplifier at 4 K and a low-noise amplifier at room temperature.  Figure \ref{fig:MW_spectrum}(b) shows the microwave spectrum for the EO converter chip, and (c) gives a zoomed-in view of the microwave mode used for EO conversion.

The intrinsic Q factors for all the devices measured here ranged from about 600 to 6000.  In similar CPW devices that were coated with the same EO polymer but fabricated on a high-resistivity silicon substrate, we measured intrinsic Q's between 30,000 and 40,000.  This leads us to believe that the Q's of these devices are not currently limited by the EO polymer. One possibility is that the Q's are limited instead by the SOI substrate.  The SOI used here is float-zone SOI from Shin-Etsu, with a large substrate resistivity $>3~ {k}\Omega\cdot$cm. However, it has been shown by Wu et al. in \cite{Wu1999} that a low resistivity inversion layer forms at the interface between high-resistivity silicon and silicon oxide, and that this layer can lead to microwave losses greater than 10 dB/cm in coplanar waveguides at gigahertz frequencies.  If this is indeed the main source of loss in the resonators, then it should be possible to increase the microwave Q of future devices by selectively removing the silicon oxide and fabricating the CPW directly on the high-resistivity silicon substrate, while keeping still keeping the oxide in the vicinity of the optical waveguides.  This could be achieved using a masked hydrofluoric acid etch.

\section{Microwave-to-optical conversion}

\begin{figure} [htbp]
\centering
\includegraphics[width=5 in]{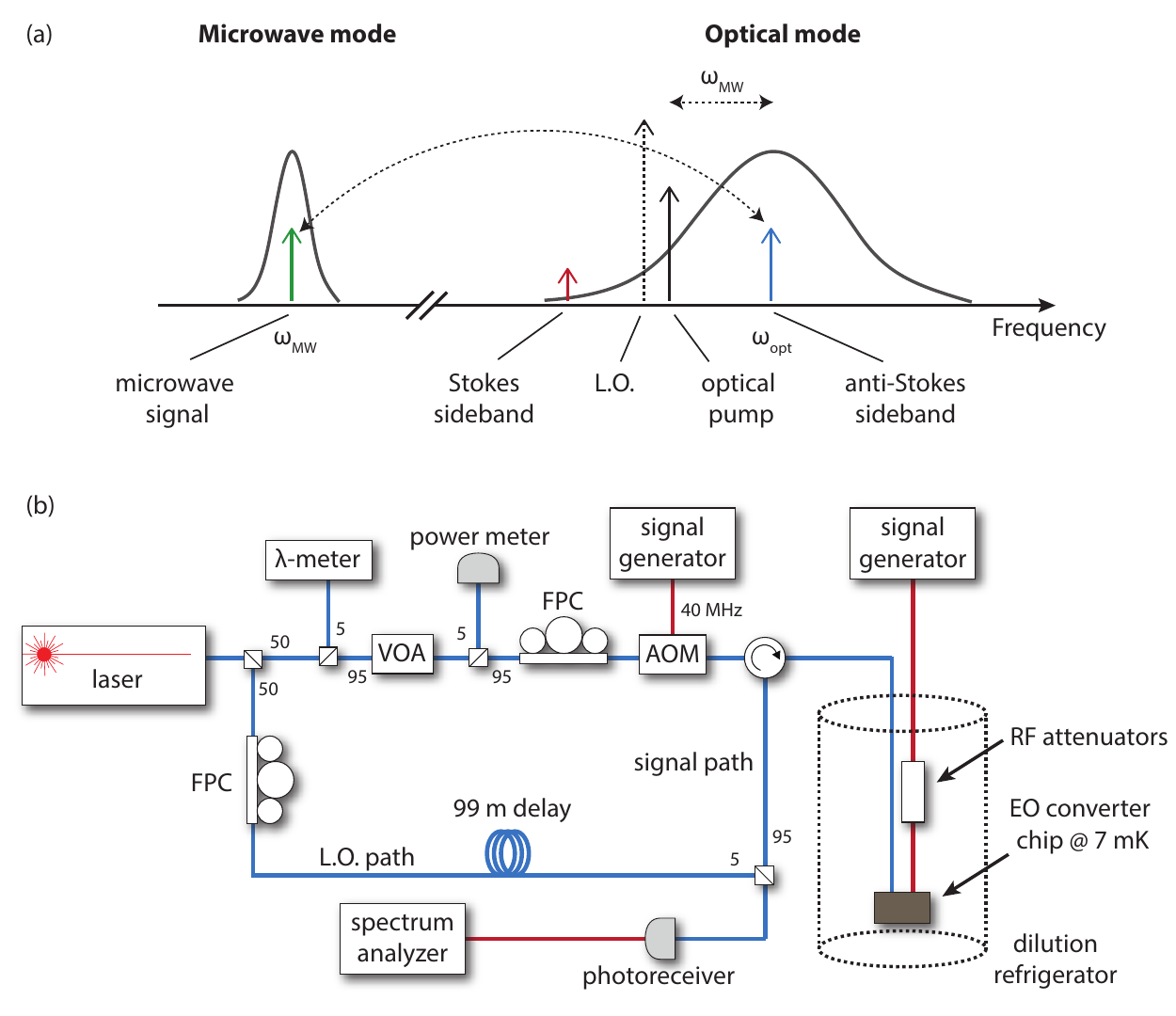}
\caption{  (a) A schematic of the conversion process for the case of red-side pump detuning.  In this mode, the device implements a beamsplitter Hamiltonian which coherently converts microwave photons to optical anti-Stokes sideband photons.  We directly measure the power in the Stokes and anti-Stokes sidebands by beating them with a local oscillator (LO) on a photoreceiver. (b) A schematic of the heterodyne measurement setup.  The blue lines represents SMF-28 optical fiber while the red lines represent electrical connections. The light from the tunable laser (Santec TSL-550) is first split with a 50/50 beamsplitter into a pump tone and (LO) tone.  On the pump path, the light is passed through an AOM which shifts the light frequency by +40 MHz.  A wavelength meter (Bristol Model 621) is used to measure the laser wavelength and an AOM allows us to control the amount of power going into the dilution refrigerator.  After being modulated in the EO converter, the reflected light from the chip is redirected to a 95/5 beamsplitter and combined with the LO.  The optical beat signals are measured with a high-speed photoreceiver and real-time spectrum analyzer.  A 99 m long fiber delay line is used to match the path lengths of the pump and LO  to reduce the laser phase noise on the measured sideband signal.  Fiber polarization controllers (FPCs) are used to match the polarization of the pump light to our on-chip grating couplers, and to match the polarization of the LO to the pump.  The converter device is driven resonantly with an RF tone from a signal generator.}
\label{fig:heterodyne}
\end{figure}

After separately outlining the device optical and microwave performance, we now turn to describe our measurement of microwave-to-optical photon conversion in a dilution refrigerator environment.  One of the key characteristics that distinguishes our electro-optic converter from a standard electro-optic modulator is that the optical resonator in our device allows us to selectively generate either the red or blue optical sideband depending on the detuning of the optical pump tone with respect to the optical cavity.  An illustration of the conversion process for the case of a red-detuned pump is shown in Figure \ref{fig:heterodyne}(a).  In this case, the device implements a beamsplitter Hamiltonian which coherently converts microwave photons to optical anti-Stokes sideband photons.  

In order to directly demonstrate the sideband selectivity of the conversion process, we use a heterodyne measurement technique illustrated in Figure \ref{fig:heterodyne}(b).  The laser light is first split into a pump tone and a local oscillator (LO) tone. The pump light is sent through an acousto-optic modulator (AOM) which shifts the frequency of the light by +40 MHz.  The pump light is then sent to the converter device inside the dilution refrigerator. The chip is driven on-resonance by an RF signal generator (Keysight E8257D), which populates the CPW resonator with microwave photons and produces a time-varying electric field on the capacitor.  This modulation produces sidebands on the light in the optical cavity, which is then reflected out of the chip and combined with the LO on a high-speed photoreceiver (New Focus 1554-B).  The resulting RF signal is sent to a spectrum analyzer (Rohde \& Schwarz FSW) to measure the power in the different sidebands.

We follow several steps to calculate and calibrate the device efficiency from the RF spectrum obtained from the heterodyne measurement. For each measurement, we subtract dark background data taken with no input signal from the signal data. The difference in power spectral density between the signal data and dark data (measured adjacent to the spectral peak of interest) gives the shot noise level, which is used for calibration. After subtracting the dark background and shot noise level from the signal spectrum, we integrate the peak of interest to find the total power.  We then divide the noise-subtracted integrated power by the adjacent shot noise level to convert a given peak power to optical photon flux (units of photons/second).
Finally, efficiency is calculated by dividing the optical sideband photon flux generated by the microwave photon flux incident on the converter. 

\begin{figure} [htbp]
\centering
\includegraphics[width=\textwidth]{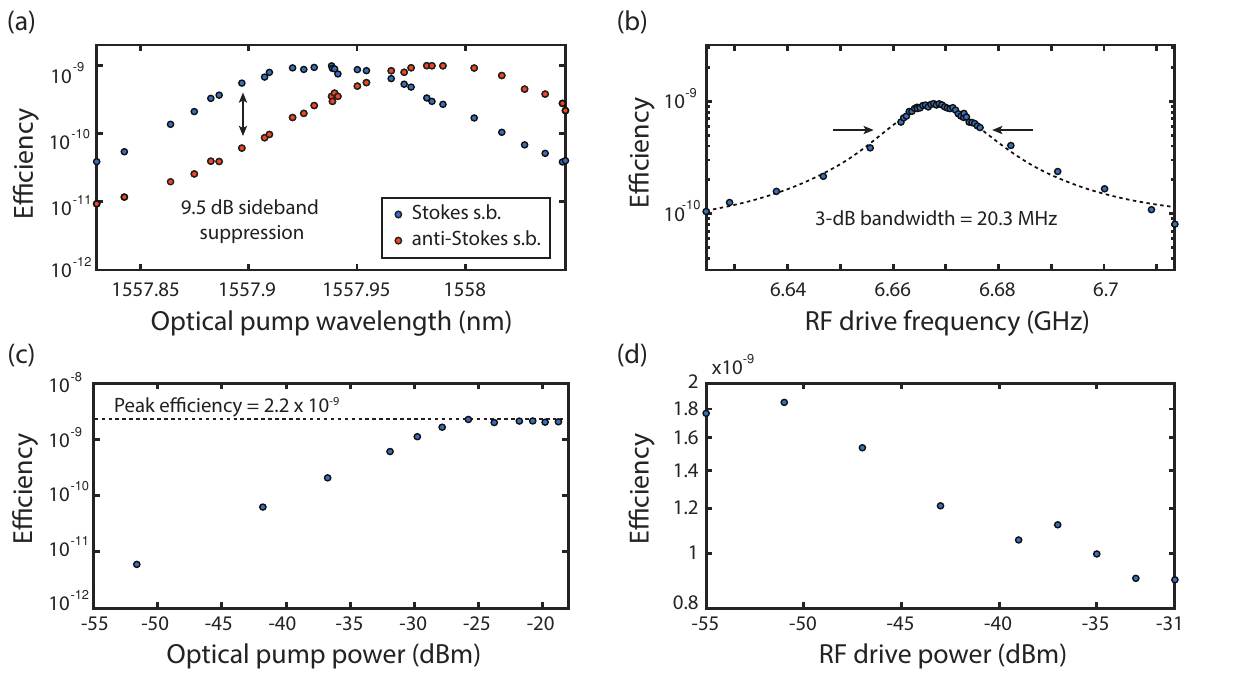}
\caption{(a) A plot of the microwave-to-optical photon conversion efficiency vs. optical pump wavelength.  A red-detuned pump selectively enhances the Stokes sideband (s.b.) while a blue-detuned pump enhances the anti-Stokes sideband. (b)   A plot of conversion efficiency vs. RF drive frequency.  The dashed line is a Lorentzian fit to the data, and it is used to extract a conversion bandwidth of 20.3 MHz.  (c) A plot of conversion efficiency vs. optical pump power (measured at the input of the dilution refrigerator).  We find that the conversion efficiency saturates at a maximum value of $2.2 \pm 0.7 \times 10^{-9}$ at an optical power of $-26$ dBm, due to the decrease of microwave Q caused by stray light.  (d) A plot of conversion efficiency vs. the estimated RF drive power in the on-chip microwave feedline.  The gradual decrease in efficiency with increasing drive power is likely due to microwave frequency absorption in the resonator, which leads to a decreased Q factor.  For all plots, the nominal experimental parameters (when not otherwise being swept) are $P_\textrm{\scriptsize pump} = -30$ dBm and $P_\textrm{\scriptsize RF} = -31$ dBm,  with the optical pump wavelength $\lambda_\textrm{\scriptsize pump} \approx  1557.92$ nm (red-detuned from the resonance by approximately $\omega_\textrm{\scriptsize MW}$) and the RF drive on-resonance with the microwave resonator.  In (b), (c) and (d) the plotted efficiency is for conversion into the Stokes sideband.  The calibration procedure to extract the microwave-to-optical conversion efficiency is described in the text.}
\label{fig:conversion_efficiency}
\end{figure}

Figure \ref{fig:conversion_efficiency} shows the microwave-to-optical conversion efficiency as a function of different experimental parameters.  Although our device is not strictly in the sideband-resolved regime ($\kappa_\textrm{\scriptsize tot}$ is slightly larger than $\omega_\textrm{\scriptsize MW}$ due to the optical mode being overcoupled), the sideband selectivity of the EO conversion process can still be seen in Figure \ref{fig:conversion_efficiency}(a). When the pump laser is red-detuned (blue-detuned) from the optical resonance, the anti-Stokes (Stokes) sideband becomes enhanced. The maximum contrast we observe between the two sidebands is 9.5 dB.  Figure \ref{fig:conversion_efficiency}(b) shows how the conversion efficiency changes as a function of microwave drive frequency.  From this data, we can directly read off the full-width-half-max conversion bandwidth of 20.3 MHz.  The effects of increasing optical pump power can be seen in Figure \ref{fig:conversion_efficiency}(c).  Initially, we see the efficiency increasing as the intracavity photon number increases.  However, the efficiency eventually saturates for optical pump powers greater than $-20$ dBm due to a trade-off between increasing intracavity photon number and decreasing microwave quality factor.  The impact of optical pump light on the microwave resonator is described in more detail in Section \ref{sec:quasiparticles}.  Figure \ref{fig:conversion_efficiency}(d) shows the conversion efficiency as a function of microwave drive power.  Although the effect is small, we do see a slight decrease in efficiency as the microwave power is increased.  We attribute this to microwave absorption in the superconducting resonators, which causes increased quasiparticle density and decreases the Q factor \cite{Mattis1958,Visser}.  

Overall, the peak conversion efficiency that we observe is $2.2 \pm 0.7 \times 10^{-9}$, with an optical pump power of $-26$ dBm and microwave input power of $-31$ dBm.  This efficiency is quoted immediately external to the packaged EO converter chip; it includes the loss of the optical grating coupler but not the loss of the upstream RF lines or the downstream optical fiber components.
The uncertainty in the efficiency is systematic and is due primarily to our uncertainty in estimating the loss of the RF input lines and optical fibers in the dilution fridge.  Based on the physics of the device, we expect the conversion efficiency for the reverse conversion process (optical photon converted to microwave photon) to be the same as the forward direction (microwave-to-optical).  However, because of the low efficiency of the device we were not able to measure the optical-to-microwave conversion efficiency directly.

To infer the electro-optic coupling rate $g_0$ from the measured conversion efficiency $\eta$, we use 
\begin{equation}
    g_0 \approx \frac{\kappa_\textrm{\scriptsize tot} \gamma_\textrm{\scriptsize tot}}{4 \sqrt{n_\textrm{\scriptsize cav}}} \sqrt{\frac{\eta}{\kappa_e \gamma_e}}
    \label{eqn:calibrate_g0}
\end{equation}
which is valid in the low cooperativity regime $C \ll 1$. 
We estimate that the $g_0$ achieved in our system is $330 \pm 60$ Hz, where the uncertainty comes from our calibration of the conversion efficiency.  This is only slightly lower than the value of 400 Hz that we would predict based on the room temperature measurements of resonator tuning.  A summary of the nominal device parameters is given in Table \ref{tab:parameters}.

\begin{table}[htbp]
\centering
\begin{tabular}{|c|c|c|}
\hline
Device parameter  & Description  & Value \\ \hhline{|=|=|=|}
$\omega_\textrm{\scriptsize opt}/2\pi$ &  optical resonance frequency &   192.6 THz  \\ \hline
$\kappa_i/2\pi$ & intrinsic optical loss rate & 2.07 GHz\\ \hline
$\kappa_e/2\pi$ &  extrinsic optical loss rate &    7.61 GHz  \\ \hline
$\kappa_\textrm{\scriptsize tot}/2\pi$ &  total optical loss rate &    9.68 GHz  \\ \hline
$\omega_\textrm{\scriptsize MW}/2\pi$ & microwave resonance frequency &    6.672 GHz   \\ \hline
$\gamma_i/2\pi$ &  intrinsic microwave loss rate   &   2.53 MHz \\ \hline
$\gamma_e/2\pi$ & extrinsic microwave loss rate  &  1.91 MHz    \\ \hline
$\gamma_\textrm{\scriptsize tot}/2\pi$ & total microwave loss rate  &  6.35 MHz    \\ \hline
$g_0/2\pi$ &  electro-optic coupling rate  &  400 Hz   \\
 & (predicted from room temperature $g_\textrm{\scriptsize V}$) & \\ \hline
$g_0/2\pi$ &  electro-optic coupling rate  &  330 Hz   \\
 & (inferred from conversion efficiency) & \\ \hline
\end{tabular}
\caption{Summary of measured device parameters.  Note that the optical resonator has single-sided coupling, so  $\kappa_\textrm{\scriptsize tot} = \kappa_i + \kappa_e$, while the microwave resonator has two-sided coupling, so $\gamma_\textrm{\scriptsize tot} = \gamma_i + 2\gamma_e$}.
\label{tab:parameters}
\end{table}

\section{Stray light and quasiparticle effects}
\label{sec:quasiparticles}
Many approaches for performing microwave-to-optical transduction utilize superconducting resonators in close proximity to intense optical light \cite{Andrews2014a,Dieterle2016,Fan2018,Soltani2017,Rochman2019}.  Since optical photons are well above the superconducting gap energy, stray light impinging on the superconductor will break Cooper pairs and excite quasiparticles \cite{Owen1972}, thereby adding excess loss to the microwave resonator and degrading the microwave-to-optical conversion efficiency.  It is therefore important to understand the effects of stray light and quasiparticle generation in these devices.  Fortunately, the microwave kinetic inductance detector (MKID) community has been studying the effects of quasiparticle generation in microwave resonators for several decades \cite{Zmuidzinas2012}, and we are able to draw from that literature here.

\subsection{Static quasiparticle effects}
\label{sec:static_QPs}
In analyzing the effect of optically generated quasiparticles on superconducting resonators, we follow the readable analyses of de Visser \cite{Visser} and Zmuidzinas \cite{Zmuidzinas2012}. From Mattis-Bardeen theory for BCS superconductors, we know that a finite temperature quasiparticle bath causes a superconductor to manifest a frequency-dependent complex conductivity, $\sigma(\omega,T) = \sigma_1(\omega,T) + i\sigma_2(\omega,T)$.  Specifically, for microwave excitation frequency $\omega$ below the superconducting gap, the components of the conductivity are given by \cite{Mattis1958}
\begin{equation}
    \frac{\sigma_1(\omega,T)}{\sigma_n} = \frac{2}{\hbar\omega} \int_\Delta^\infty ~dE  \frac{E^2 + \Delta^2 + \hbar\omega E}{\sqrt{(E^2 - \Delta^2)}\sqrt{(E+\hbar\omega)^2-\Delta^2}} \left[ f(E) - f(E+\hbar\omega)\right]~~~~
\end{equation}
\begin{equation}
    \frac{\sigma_2(\omega,T)}{\sigma_n} = \frac{1}{\hbar\omega} \int_{\Delta-\hbar\omega}^\Delta ~dE \frac{E^2 + \Delta^2 + \hbar\omega E}{\sqrt{( \Delta^2 - E^2)}\sqrt{(E+\hbar\omega)^2-\Delta^2}} \left[ 1 - 2f(E+\hbar\omega)\right]~~~~
\end{equation}
 where $\Delta$ is the superconducting gap, and $\sigma_n$ is the normal state conductivity. $f(E)$ is the quasiparticle distribution function which in the case of thermal equilibrium is equal to the Fermi-Dirac distribution $ f(E) = 1/(1+\exp{(E/k_B T)})$. It should be stressed that the relevant temperature here is the temperature of the quasiparticle bath, which in the presence of stray light can be much higher than the base temperature of the cryostat. For quasiparticle temperatures greater than about $T_c/4$, it becomes necessary to take into account the temperature dependence of the superconducting gap $\Delta$, which is given implicitly by
 \begin{equation}
     \frac{1}{N_0 V_\textrm{\scriptsize sc}} = \int_{\Delta(T)}^{k_B T_D} ~dE \frac{1-2f(E)}{\sqrt{E^2 - \Delta^2(T)}}
 \end{equation}
 where $N_0$ is the single-spin density of electron states at the Fermi energy, $V_\textrm{\scriptsize sc}$ is the potential energy for electron-phonon exchange, and $T_D$ is the Debye temperature for the material.  
 
 The next step is to translate this into a measurable effect on our resonators.  The complex conductivity leads to a complex surface impedance, which in the so-called dirty limit (where the electron mean free path is less than the Cooper pair coherence length) is given by \cite{Visser}
\begin{equation}
    Z_s = R_s + i\omega L_s = \sqrt{\frac{i\mu_0\omega}{\sigma_1-i\sigma_2}}\coth{\left(d\sqrt{i\omega\mu_0}\right)},
    \label{eqn:Z_s}
\end{equation}
where $\mu_0$ is the permeability of free space and $d$ is the superconductor film thickness. We call $R_s$ the sheet resistance and $L_s$ the sheet kinetic inductance.

The impact of the sheet kinetic inductance on a microwave resonator depends on the exact resonator geometry.  For a CPW geometry the kinetic inductance per unit length can be calculated as $L_k = (g_c + g_g)L_s$, where $g_c$ and $g_g$ are factors accounting for the contributions of the center conductor and groundplane, respectively.  The formulae for $g_c$ and $g_g$ are given in \cite{Visser}.  The kinetic inductance adds to the bare geometric inductance of a quarter-wave CPW resonator so that its resonance frequency is given by $\omega_0 = \frac{2\pi}{4l\sqrt{L_\textrm{\scriptsize tot}C}}$, where $l$ is the CPW length, $L_\textrm{\scriptsize tot} = L_\textrm{\scriptsize geom} + L_k$ is the total inductance per unit length and $C$ is the capacitance per unit length. A useful quantity to keep in mind is the kinetic inductance fraction, defined as $\alpha_k = \frac{L_k}{L_\textrm{\scriptsize tot}}$, because it gives a measure of how sensitive a resonator geometry is to changes in quasiparticle density.   We estimate that our CPW resonators have an $\alpha_k$ of approximately 5\% (for $T \ll T_c$).  The sheet resistance $R_s$ in Equation \ref{eqn:Z_s} has the effect of reducing the quality factor of a resonator.  The quasiparticle-limited Q factor is given by $Q_\textrm{\scriptsize qp} = \frac{\omega L}{R} = \frac{1}{\alpha_k}\frac{\omega L_s}{R_s}$.

To summarize, the preceding analysis shows that given a quasiparticle bath temperature $T$, we can calculate the shift in resonator frequency and quality factor, and vice versa.  From the bath temperature we can also calculate the quasiparticle density as \cite{Visser}
\begin{equation}
    n_\textrm{\scriptsize qp} = 4 N_0\int_\Delta^\infty ~dE \frac{E}{\sqrt{E^2 - \Delta^2}}f(E).
\end{equation}
The relevant superconducting parameters for our aluminum films are summarized in Table \ref{tab:quasiparticles}.

\begin{figure} [htbp]
\centering
\includegraphics[width=5 in]{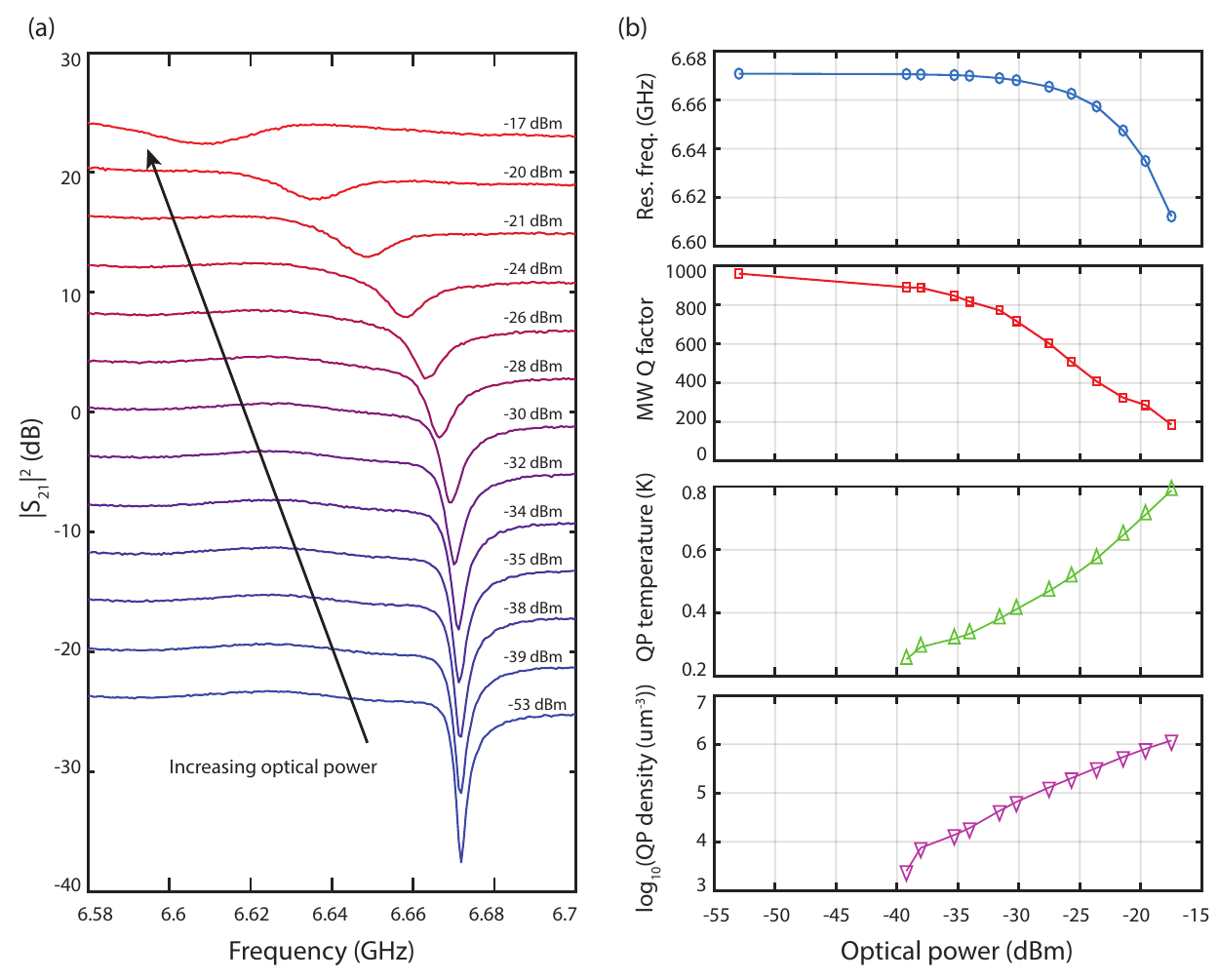}
\caption{(a) Spectra of the microwave resonance, plotted for increasing CW optical power levels (labeled for each curve).  The vertical axis is referenced to the bottom spectrum ($-53$ dBm optical power), with each subsequent spectrum offset by +4 dB for clarity.  (b) The resonance from each spectrum in (a) is fit with a Fano-Lorentzian lineshape and the extracted resonance frequency and total Q factor are plotted vs. optical power.  The quasiparticle bath temperature and quasiparticle density are calculated from the change in resonance frequency using the procedure described in the text.  The lines connecting points in (b) are meant as guides to the eye.}
\label{fig:quasiparticles}
\end{figure}

To investigate the effects of quasiparticle generation in our device, we took microwave spectra of the resonator described in Section \ref{sec:microwave} while sending different CW optical power levels to the optical cavity.  Figure \ref{fig:quasiparticles}(a) shows that the resonance frequency and Q factor both decrease as the optical power is increased, as expected. Using the analysis described above, we can also convert the measured change in resonance frequency to extract the quasiparticle bath temperature and density as a function of optical power (Figure \ref{fig:quasiparticles}(b)).  Importantly, we observe a significant decrease in the microwave Q factor even for modest optical pump powers of 1 $\mu$W ($-30$ dBm). For comparison, the authors in \cite{Fan2018} used a NbTiN resonator at 2 K and were able to use a large optical pump power of 14 dBm without significant adverse effects.  The reduced susceptibility of that device to stray light is likely due to the higher $T_c$ and shorter quasiparticle lifetime of NbTiN compared to Al (discussed more in Section \ref{sec:reducing_stray_light}).  Additionally, the chip in that experiment was submersed in liquid helium to provide greater cooling power, while the present experiment relied on the cooling power of the dilution fridge mixing chamber, which is only about 4 $\mu$W at 10 mK (much smaller than typical cryostats operating in the 1--4 K range).

It is instructive to compare the estimated quasiparticle bath temperature with the temperature of the dilution fridge mixing chamber, as measured using a resistive thermometer.  For the largest optical power of $-17$ dBm (20 $\mu$W) we observed the mixing chamber steady-state temperature rise from $\approx 7$ mK to 18 mK, while for lower optical powers there was no measured change.  Clearly this small temperature change cannot account for the estimated quasiparticle temperature of $\approx 0.8$ K, and so this suggests that the quasiparticle generation we are observing is due to local absorption on the chip rather than a large-scale heating of the entire mixing chamber.

\begin{table}[htbp]
\centering
\begin{tabular}{|c|c|c|}
\hline Parameter & Description  & Value \\ \hhline{|=|=|=|}
$\sigma_n$ &  normal state conductivity &  $1.3\times 10^8 ~\Omega^{-1}{m}^{-1}$   \\ \hline
$T_c$ & superconducting critical temperature & 1.1 K\\ \hline
$\Delta_0$ & superconducting gap at zero-temperature & 167 $\mu$eV \\ \hline
$N_0$ & single-spin density of states at Fermi energy & $1.72 \times 10^{10} ~eV^{-1} \mu {m}^{-3}$ \\ \hline
$d$ & metal film thickness & 100 nm \\ \hline
$L_s$ & sheet inductance at $T = 0$ K & 140 fH/square \\ \hline
$\tau_0$ & electron-phonon interaction time & $458$ ns \\ \hline
\end{tabular}
\caption{Parameters for superconducting aluminum films used for quasiparticle calculations.  $\tau_0$ is taken from \cite{Visser}.}
\label{tab:quasiparticles}
\end{table}

\subsection{Pulsed operation}

One possibility for avoiding the effects of heating and quasiparticle generation in an electro-optic converter is to operate the device in a pulsed mode.  In this scheme, the optical pump light would be switched on for a brief interval, during which the device could perform microwave-to-optical conversion with some efficiency.  If this period was short enough, the conversion could finish before the microwave Q factor became degraded---that is, before the quasiparticle population came to thermal equilibrium with the rest of the system---and therefore achieve a higher peak efficiency.  By keeping the duty cycle low, it would be possible to decrease the steady-state heat load on the device, albeit at the expense of conversion throughput.

To investigate the feasibility of pulsed operation in our system,  we shine pulsed laser light onto the device and measure the time-dependent response of the microwave resonator. Experimentally, we implement the pulsing by switching on and off the RF power to the acousto-optic modulator shown in Figure \ref{fig:heterodyne}(a).  This allows the AOM to act as a fast optical switch with 53 dB extinction and a rise-time of about 100 ns.  While the light is being pulsed, we excite the resonator with a CW RF tone from a signal generator, and measure the time-dependent magnitude and phase of the transmitted RF signal using the real-time spectrum analyzer in IQ mode.  By  stepping the RF excitation frequency on the signal generator we can build up a time-dependent microwave spectrum.  

\begin{figure} [htbp]
\centering
\includegraphics[width=5 in]{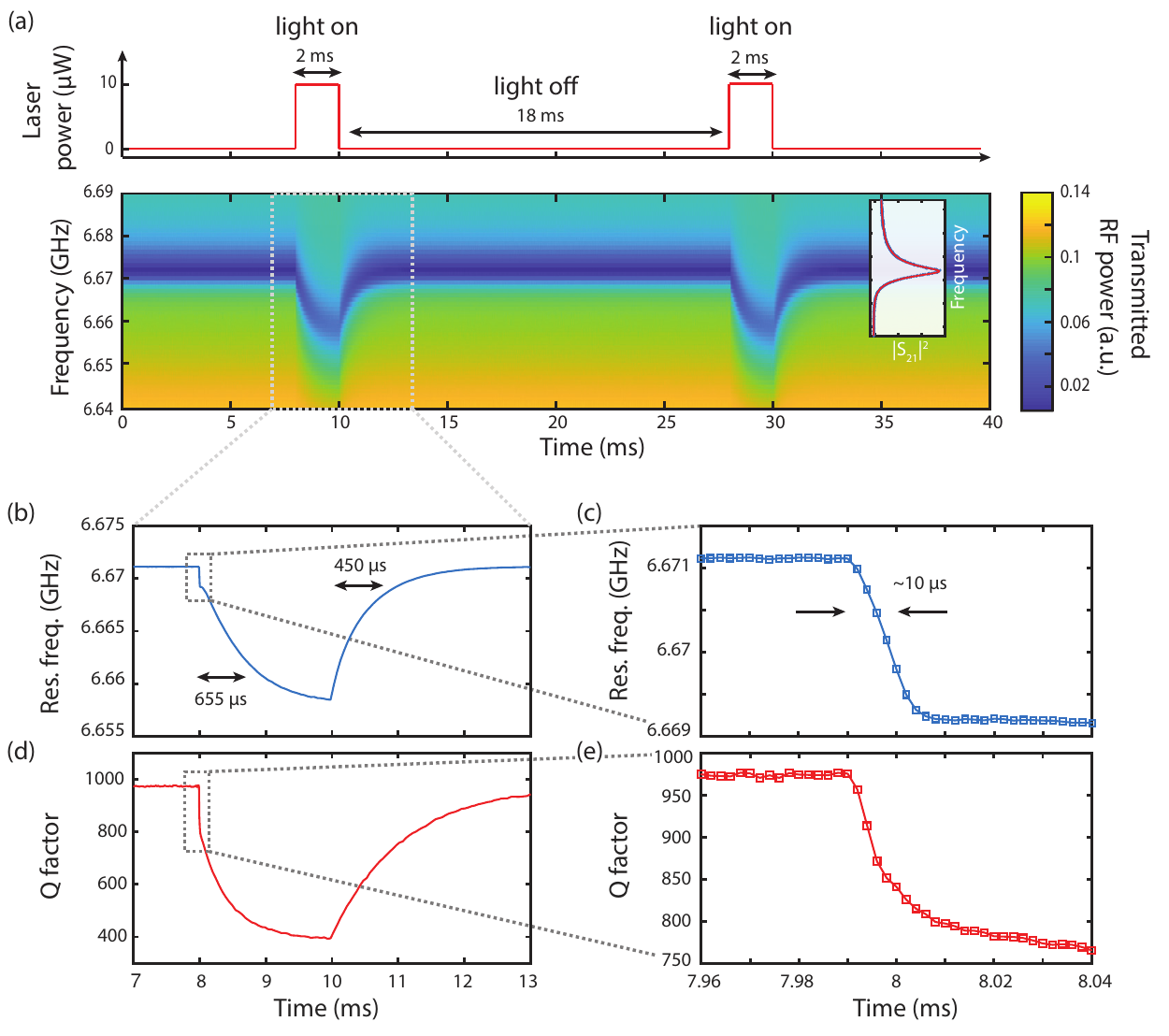}
\caption{(a) The color plot shows a time-dependent spectrum of the microwave resonator, illustrating the effects of pulsed light on the microwave mode.  The 2 ms long light pulse is switched on at approximately $t = 8$ ms and repeated every 20 ms. The color plot is constructed by vertically stacking individual time traces (horizontal lines), each taken at a different frequency.  Each vertical slice of the color plot gives the microwave spectrum at a particular point in time (as shown in the inset).  (b) and (d): Vertical slices from the color plot in (a) are selected out and fit to extract the resonance frequency and total Q factor as a function of time. Fitting the resonance frequency changes to an exponential decay gives a rise time (fall time) of 655 $\mu$s (450 $\mu$s).  (c) and (e): Zooming-in on the initial region of (b) and (d) shows a rapid initial decrease in the resonance frequency (decrease in Q) which takes place in approximately 10 $\mu$s.}
\label{fig:pulsed}
\end{figure}

An example of such a spectrum is shown in Figure \ref{fig:pulsed}(a). The light is pulsed on for 2 ms with a period of 20 ms (10\% duty cycle).  When the light is turned on, the microwave resonator frequency and quality factor both decrease quickly, and when the light is turned off the mode quickly recovers to its original state.  Fitting the frequency change to a decaying exponential, we extract time constants of 655 $\mu$s and 450 $\mu$s for the fall and rise respectively.  If we zoom in on the initial 100 $\mu$s window when the light is first turned on, we can also see an initial sharp decrease in the resonance frequency with a much faster timescale of $\approx$ 10 $\mu$s (Figure \ref{fig:pulsed}(c)).  This initial drop is well resolved by our measurement technique and is much slower than either the AOM risetime ($\approx$ 100 ns) or the microwave resonator lifetime ($\approx$ 160 ns).  

To understand the time-scales present in the system, we first note that a pair of quasiparticles relaxes by recombining and emitting a phonon, which must then scatter out of the superconducting film \cite{Rothwarf1967}.  The quasiparticle lifetime $\tau_\textrm{\scriptsize qp}$ is theoretically predicted to follow a $1/n_\textrm{\scriptsize qp}$ dependence, given by \cite{Kaplan1976, Visser}
\begin{equation}
    \tau_\textrm{\scriptsize qp} 
    \approx \frac{\tau_0}{n_\textrm{\scriptsize qp}}\frac{N_0 (k_B T_c)^3}{2 \Delta^2},
    \label{eqn:lifetimes}
\end{equation}
where $\tau_0$ is the material dependent electron-phonon interaction time.  The approximation in Equation \ref{eqn:lifetimes} is valid for thermally distributed quasiparticles near the gap edge ($E\approx \Delta)$, with $T \ll T_c$. Experimentally the quasiparticle lifetime is observed to saturate at some maximum value, so this relationship is only accurate for sufficiently large quasiparticle density ($\gtrsim$ 100 $\mu$m$^{-3}$ for aluminum) \cite{Zmuidzinas2012}.  During the initial fast dynamics shown in Figure \ref{fig:pulsed} (c) and (e), we calculate that the quasiparticle density increases from approximately $5\times 10^3$ to $5\times 10^4$ $\mu$m$^{-3}$, which from Equation \ref{eqn:lifetimes} corresponds to a predicted decrease in the quasiparticle lifetime from roughly 30 $\mu$s to 3 $\mu$s.  Since the observed fall-time of $\approx$ 10 $\mu$s falls in this range, it seems reasonable to conclude that the time-scale of the fast dynamics here is likely set by the quasiparticle relaxation lifetime.

In contrast, the longer time-scales of 655 and 450 $\mu$s seem too long to correspond to quasiparticle lifetimes based on the quasiparticle densities inferred here. More likely these time-scales are set by thermal time constants related to heat transfer between the silicon chip and the copper PCB.  These thermal time constants are challenging to estimate because of the difficulty in finding reliable data for the heat capacity and thermal conductivity of our materials at millikelvin temperatures. We note that we have observed similar time scales (100's of microseconds) for niobium resonators on sapphire substrates at 1 K.  In that case, the niobium quasiparticle lifetime is expected to be extremely short ($<$ 1 ns, see Table \ref{tab:superconductors}) and is therefore unlikely to be involved, suggesting that thermal transport is the more likely cause.  Results from those niobium resonators will be presented in a future publication.

Finally, we note that there is a discrepancy between the maximum frequency shift of $-13$ MHz observed with pulsed light and the steady-state frequency shift of $-33$ MHz observed in Figure \ref{fig:quasiparticles}(b) for the same optical power of $-20$ dBm. This suggests that there may be even slower dynamics at play, perhaps associated with heating of the entire mixing chamber plate.  We have directly observed longer heating time scales of 1--2 seconds in other chips (niobium resonators on sapphire at 1 K), but did not attempt to look for those long time scales with the present devices.

Returning to our original motivation of pulsed converter operation, we see from Figure \ref{fig:quasiparticles}(d) and (e) that to avoid a decrease in the microwave Q factor, the conversion must take place very quickly, within a few microseconds of turning on the pump light.  This window is very short, but still several times longer than the microwave resonator lifetime, suggesting that operating in such a mode could be possible.  However, on the whole it seems preferable to reduce the impact of stray light rather than working around it, and that is the focus of the next section.

\subsection{Reducing the impact of stray light}
\label{sec:reducing_stray_light}
The effect of stray light on the microwave quality factor is one of the key limiting factors of the conversion efficiency in this demonstration.  Even using extremely limited optical pump powers of $-20$ dBm, we already see the conversion efficiency saturate due to increased microwave loss.  If instead we could use an optical pump power of 14 dBm as in \cite{Fan2018}, we anticipate the conversion efficiency to immediately increase by 3 orders of magnitude.

We can break down strategies for reducing the effect of stray light into three main approaches.  The first possibility is to directly reduce the rate of optical generation of quasiparticles.  Of course, the best approach is to reduce the source of optical scattering in the first place, e.g. by improving fiber-to-chip coupling efficiency or by reducing intrinsic loss in the optical cavity, since this is the only way to reduce steady-state heating of the dilution fridge.  Other strategies are physically separating or shielding the microwave resonator from the light or adding optically absorptive materials around the device \cite{Barends2011}.  

A second possibility is to keep the quasiparticle generation rate fixed, but to reduce the resulting quasiparticle \emph{density}.  This can be achieved by fabricating the microwave resonator from a superconductor with a short maximum quasiparticle lifetime, since a short lifetime results in a smaller steady state quasiparticle population \cite{Rothwarf1967}.  Another option is to increase the metal film thickness, since to first order this will increase the total superconductor volume without increasing the absorbed optical power, and therefore dilute the quasiparticle density.  Another potential option is adding normal metal quasiparticle traps to the resonator \cite{Lang2003}.

The third possibility is to keep the quasiparticle density fixed, but to reduce its impact on the microwave quality factor by reducing the kinetic inductance fraction.  As explained in Section \ref{sec:static_QPs}, the amount of loss added by the excess quasiparticles is directly proportional to the kinetic inductance fraction $\alpha_k$ \cite{Visser}.  As a result, loss can be minimized by designing the resonator to have a large geometric inductance, and by reducing the kinetic inductance, for example by fabricating conductors that are wide and thick and by using superconductors with small penetration depths.

The choice of superconductor is clearly important.  To reduce the impact of optically generated quasiparticles we want to choose a superconductor that has a short quasiparticle lifetime and a small penetration depth $\lambda_0$.   Interestingly, this is exactly the opposite criteria compared to what is desirable for an MKID detector, where the goal is to make a resonator as sensitive to optically generated quasiparticles as possible \cite{Visser}. Having a higher $T_c$ is also useful because it reduces the impact of environmental heating on the microwave resonators and provides additional flexibility to do experiments at different locations within the dilution refrigerator besides the mixing chamber.    Some commonly used superconductors with their relevant properties are listed in Table \ref{tab:superconductors}.  Although aluminum is a commonly used material in superconducting qubits and lends itself to straightforward fabrication processes, it is a fairly poor material for stray light handling because of its extremely long quasiparticle lifetime (up to ms).  In comparison, niobium and niobium titanium nitride both have a maximum quasiparticle lifetime which is 6 orders of magnitude shorter, as well as higher $T_c$'s.  We anticipate future electro-optic converters made of these materials to exhibit greatly improved performance in the presence of stray light.

\begin{table}[htbp]
\centering
\begin{tabular}{|c|c|c|c|}
\hline
Material        & $T_c$ (K) & $\tau_\textrm{\scriptsize qp, max}$ & $\lambda_0$ (nm)  \\ \hhline{|=|=|=|=|}
Al & 1.1 & 3.5 ms & 89\\ \hline
Nb & 9.2 & 1 ns & 45 \\ \hline
TiN & 0.7 - 4.5 & 200 $\mu$s & 500-3000 \\ \hline
NbTiN & 14.5 & 1 ns & 275 \\ \hline
\end{tabular}
\caption{Relevant parameters for some commonly used superconductors.  The values here are a selection taken from a more comprehensive table in \cite{Visser}.}
\label{tab:superconductors}
\end{table}

\section{Increasing $g_0$ using high-impedance spiral resonators}
\label{sec:spiral}

Because the electro-optic coupling rate $g_0$ scales with $\sqrt{Z} = (L/C)^{\frac{1}{4}}$, as shown in Equation \ref{eqn:g0}, it is advantageous to increase the resonator impedance by making the inductance as large as possible, while reducing the capacitance to keep the resonance frequency fixed.  There are several different approaches in the literature for achieving high-impedance, low-loss microwave resonators.  One approach is to use high-kinetic inductance nanowires made of disordered superconductors such as NbN, TiN or NbTiN \cite{Samkharadze2016,Annunziata2010a,Hazard2019,Shearrow2018}.  However, depositing high quality films from these materials is generally more challenging than evaporating aluminum, often requiring carefully tuned sputtering processes or atomic layer deposition \cite{Annunziata2010a,Shearrow2018,Yemane2017}.  A second approach is to use the kinetic inductance of an array of large Josephson junctions, where the Josephson energy $E_J$ of each individual junction is much larger than the charging energy $E_C$ \cite{Masluk2012,Bell2012}.  With these devices, careful design is required to mitigate the effects of coherent quantum phase slips.  Additionally, since both Josephson junction and nanowire-based inductors rely on kinetic inductance, they will be more sensitive to quasiparticle density, and hence to stray light, compared to primarily geometric inductors, as described in the last section.

A third approach, for which we present initial results here, is to focus on devices with a large geometric inductance, such as a planar spiral inductor \cite{Dieterle2016}.  As pointed out in \cite{Fink2016}, the capacitance of a spiral inductor is mostly determined by the area of the inductor.  By shrinking the coil pitch, and thereby increasing the number of turns while keeping area fixed, it is possible to greatly increase the spiral inductance without increasing the capacitance.  With proper design, the impedance of these devices can be in the k$\Omega$ range \cite{Fink2016}.  In this case the maximum impedance is limited by the minimum fabricable wire pitch, as well as by the self-resonance frequency of the spiral, which decreases as the number of turns is increased.  Increasing the impedance of the microwave resonator in our device from 100 $\Omega$ to 10 k$\Omega$ would increase our conversion efficiency by two orders of magnitude. 

\begin{figure} [htbp]
\centering
\includegraphics[width=5 in]{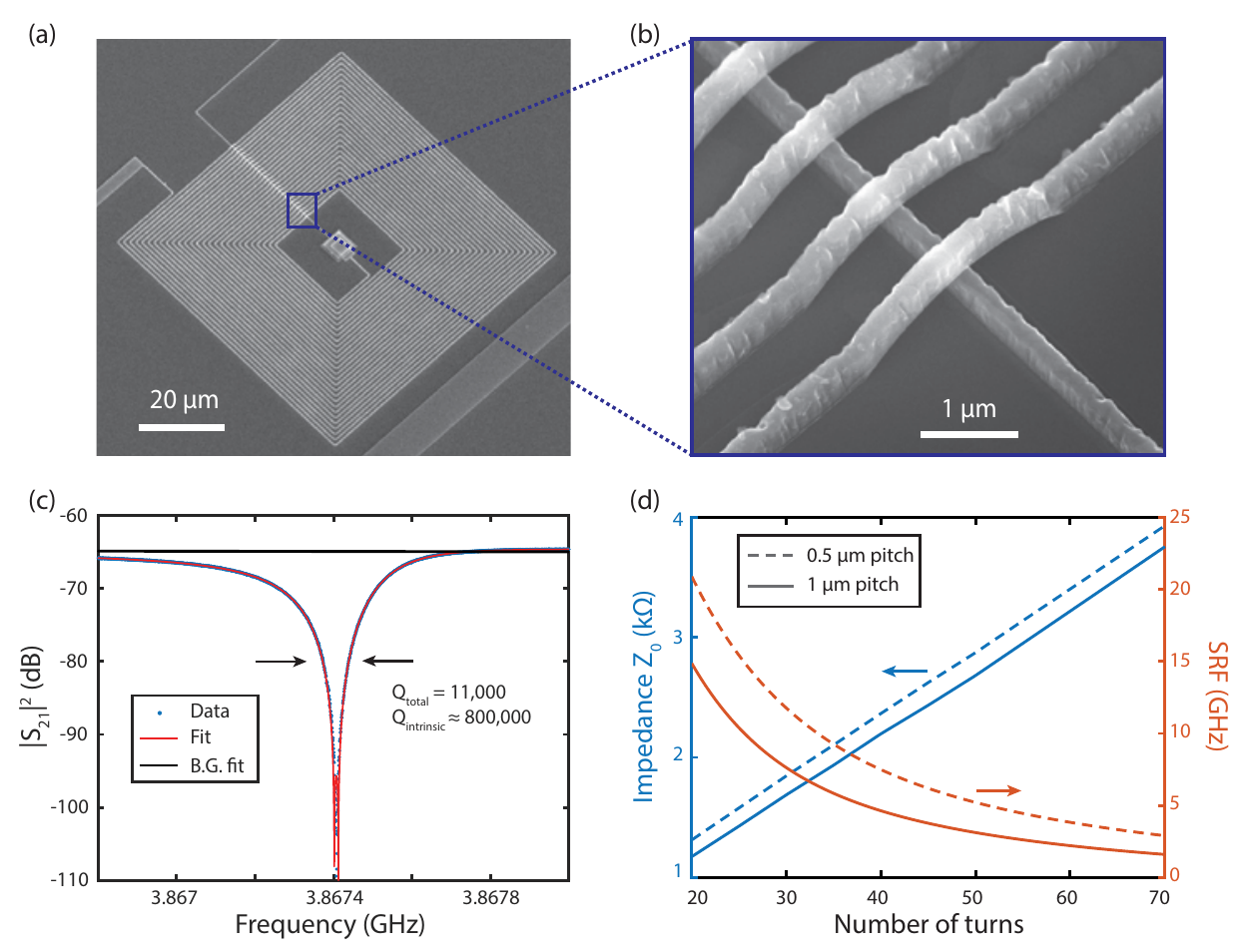}
\caption{(a) An SEM micrography of a spiral inductor with 27 turns.  The structure in the center of the spiral is a pair of overlapping pads with bandaging to ensure good electrical contact between metal layers.  (b) A zoomed-in SEM showing the aluminum airbridges jumping over the bottom contact wire. (c) A measured spectrum from an LC resonator with a spiral inductor and shunt capacitor.  Although the mode is very over-coupled, the intrinsic quality factor is estimated to be approximately 800,000. (d) Simulation results showing the impact of the number of spiral turns on the impedance and the self-resonant frequency (SRF) of a spiral inductor for wire pitches of 0.5 (dashed line) and 1 $\mu$m (solid line). The wire fill factor is 25\%. For a target self-resonant frequency, a 0.5 $\mu$m wire pitch allows for a higher impedance circuit than a 1 $\mu$m pitch.}
\label{fig:spirals}
\end{figure}

Figure \ref{fig:spirals}(d) shows simulated results of the impedance and self-resonant frequency (SRF) of a spiral inductor for two wire pitches: 0.5 $\mu$m and 1 $\mu$m. The simulated spiral sits on a silicon substrate and is coated with 500 nm of SEO100C EO polymer. The simulations were performed in Sonnet and the SRF of each spiral was found by measuring the S21 of a transmission line inductively coupled to the spiral. The value for inductance of the spiral was calculated analytically \cite{Mohan1999} and verified in Sonnet. Figure \ref{fig:spirals}(d) illustrates that an arbitrarily high impedance cannot be reached due to the SRF of the spiral which decreases as the impedance of the spiral increases. 
The total resonant frequency of the circuit will also include the electrode capacitance in parallel with the self-capacitance of the spiral, so the inductor must be sized to allow for the additional electrode capacitance.  

An example spiral inductor is shown in Figure \ref{fig:spirals}(a).  A key feature of the spiral geometry is that the wires in the coil region must be able to ``jump'' over the central feed wire.  To enable this, we use an airbridge fabrication process similar to the one described in \cite{Norte2015}.  After depositing the bottom feed wire, we use e-beam lithography to pattern scaffolds out of LOR-5B. We then heat the chip to reflow the LOR-5B, giving the scaffolds a rounded profile.  Finally, we do another e-beam lithography step and deposit the coil wire on top of the scaffolds.  The LOR-5B dissolves in the lift-off process and the aluminum wires are left with suspended airbridges over the bottom contact wire, as shown in Figure \ref{fig:spirals}(b).

The microwave spectrum from an example device consisting of a spiral inductor with a shunt capacitor is shown in Figure \ref{fig:spirals}(c).  This device had a total impedance of 1.2 k$\Omega$.
The spiral LC resonator was fabricated on the same Shin-Etsu SOI substrate previously described, but with the buried oxide and silicon device layer both etched away, and the circuit was not coated in EO polymer. The device had a large high-power intrinsic quality factor of approximately 800,000.  We believe that high-impedance resonators such as this one could prove useful for the next generation of EO conversion experiments.

\section{Conclusions and outlook}

We have presented a new platform for quantum microwave-to-optical transduction based on silicon-organic hybrid photonics.  This platform takes advantage of the large electro-optic coefficient of EO polymers and the mature nanofabrication associated with SOI technology.  Our doubly-resonant design allows the device to achieve sideband-selective photon conversion.  The device has a large bandwidth of 20.3 MHz, and the measured coupling rate $g_0/2\pi$ of 330 Hz is comparable to that of other electro-optic transduction approaches \cite{Fan2018,Rueda2016}. The device was demonstrated in a millikelvin dilution refrigerator environment, an important step towards reducing the thermal noise added during transduction and bringing us closer to interfacing with superconducting qubits.

Although the conversion efficiency of $2.2 \times 10^{-9}$ demonstrated here is very modest, we note that there are straightforward paths for improvement.  For example, as mentioned earlier, the conversion efficiency was decreased by a factor of 11 due to an avoidable packaging issue.  We also anticipate that the conversion efficiency could be improved by 2--3 orders of magnitude by changing the metal from aluminum to Nb or NbTiN to allow for larger optical pump powers, and another order of magnitude by increasing the impedance of the microwave resonator.  Since the microwave quality factor seems to be limited by the SOI substrate rather than the EO polymer, switching to a lower loss substrate such as sapphire may provide an additional order of magnitude.  One aspect which may prove more challenging to improve is the optical Q factor, since this seems to be limited directly by poling induced loss in the EO polymer. Overall, we expect that this platform has the potential to make a significant contribution to the technology of future quantum networks.

\section*{Acknowledgments}
This work was supported by the US Government through the National Science Foundation under grant No. ECCS-1708734, Army Research Office  (ARO/LPS) CQTS program, through Airforce Office of Scientific Research (AFOSR) via (MURI No. FA9550-17-1-0002 led by CUNY). Part of this work was performed at the Stanford Nano Shared Facilities (SNSF) and Stanford Nanofabrication Facility (SNF). SNSF is supported by the National Science Foundation under award ECCS-1542152. A.S.N. acknowledges the support of a David and Lucile Packard Fellowship. R.V.L. acknowledges funding from VOCATIO and from the European Union's Horizon 2020 research and innovation program under Marie Sk\l{}odowska-Curie grant agreement No. 665501 with the research foundation Flanders (FWO). J.D.W. and P.A.A. acknowledge support from a Stanford Graduate Fellowship. E.A.W. acknowledges support by the Department of Defense (DoD) through the National Defense Science \& Engineering Graduate Fellowship (NDSEG) Program.

\section*{References}

\bibliographystyle{iopart-num}

\begin{thebibliography}{10}
\expandafter\ifx\csname url\endcsname\relax
  \def\url#1{{\tt #1}}\fi
\expandafter\ifx\csname urlprefix\endcsname\relax\def\urlprefix{URL }\fi
\providecommand{\eprint}[2][]{\url{#2}}

\bibitem{Devoret2013a}
Devoret M~H and Schoelkopf R~J 2013 {\em Science (80-. ).\/} {\bf 339}
  1169--1175

\bibitem{Wendin2017}
Wendin G 2017 {\em Reports Prog. Phys.\/} {\bf 80} ISSN 00344885

\bibitem{Kimble2008a}
Kimble H~J 2008 {\em Nature\/} {\bf 453} 1023--1030 ISSN 14764687

\bibitem{Lauk2019}
Lauk N, Sinclair N, Barzanjeh S, Covey J~P, Saffman M, Spiropulu M and Simon C
  2019 Perspectives on quantum transduction  1--13 (arXiv:\eprint{1910.04821})

\bibitem{Lambert2019}
Lambert N~J, Rueda A, Sedlmeir F and Schwefel H~G~L 2019 Coherent conversion between microwave and optical photons -- an overview of physical implementations  1--17
  (arXiv:\eprint{1906.10255})

\bibitem{Andrews2014a}
Andrews R~W, Peterson R~W, Purdy T~P, Cicak K, Simmonds R~W, Regal C~A and
  Lehnert K~W 2014 {\em Nat. Phys.\/} {\bf 10} 321--326 ISSN 17452481

\bibitem{Bagci2014}
Bagci T, Simonsen A, Schmid S, Villanueva L~G, Zeuthen E, Appel J, Taylor J~M,
  S{\o}rensen A, Usami K, Schliesser A and Polzik E~S 2014 {\em Nature\/} {\bf
  507} 81--85 ISSN 14764687

\bibitem{Forsch2019a}
Forsch M, Stockill R, Wallucks A, Marinkovi{\'{c}} I, G{\"{a}}rtner C, Norte
  R~A, van Otten F, Fiore A, Srinivasan K and Gr{\"{o}}blacher S 2019 {\em Nat.
  Phys.\/} ISSN 1745-2473

\bibitem{Vainsencher2016}
Vainsencher A, Satzinger K~J, Peairs G~A and Cleland A~N 2016 {\em Appl. Phys.
  Lett.\/} {\bf 109} ISSN 00036951

\bibitem{Balram2015b}
Balram K~C, Davan{\c{c}}o M~I, Song J~D and Srinivasan K 2016 {\em Nat.
  Photonics\/} {\bf 10} 346--352 ISSN 1749-4885

\bibitem{Jiang2019}
Jiang W, Sarabalis C~J, Dahmani Y~D, Patel R~N, Mayor F~M, McKenna T~P, {Van
  Laer} R and Safavi-Naeini A~H 2019 Efficient bidirectional piezo-optomechanical transduction between microwave and optical frequency (arXiv:\eprint{1909.04627})

\bibitem{Shao2019}
Shao L, Yu M, Maity S, Sinclair N, Zheng L, Chia C, Shams-Ansari A, Wang C,
  Zhang M, Lai K and Loncar M 2019 Microwave-to-optical conversion using lithium niobate thin-film acoustic resonators (arXiv:\eprint{1907.08593})

\bibitem{Rueda2016}
Rueda A, Sedlmeir F, Collodo M~C, Vogl U, Stiller B, Schunk G, Strekalov D~V,
  Marquardt C, Fink J~M, Painter O, Leuchs G and Schwefel H~G~L 2016 {\em
  Optica\/} {\bf 3} 597 ISSN 2334-2536

\bibitem{Fan2018}
Fan L, Zou C~L, Cheng R, Guo X, Han X, Gong Z, Wang S and Tang H~X 2018 {\em
  Sci. Adv.\/} {\bf 4} 1--6 ISSN 23752548

\bibitem{Hisatomi2016}
Hisatomi R, Osada A, Tabuchi Y, Ishikawa T, Noguchi A, Yamazaki R, Usami K and
  Nakamura Y 2016 {\em Phys. Rev. B - Condens. Matter Mater. Phys.\/} {\bf 93}
  174427 ISSN 1550235X

\bibitem{Vogt2019}
Vogt T, Gross C, Han J, Pal S~B, Lam M, Kiffner M and Li W 2019 {\em Phys. Rev.
  A\/} {\bf 99} 1--6 ISSN 24699934

\bibitem{Han2018}
Han J, Vogt T, Gross C, Jaksch D, Kiffner M and Li W 2018 {\em Phys. Rev.
  Lett.\/} {\bf 120} 93201 ISSN 10797114

\bibitem{Fernandez-gonzalvo2019}
Fernandez-Gonzalvo X, Horvath S~P, Chen Y~h and Longdell J~J 2019 {\em Phys.
  Rev. A\/} {\bf 100}

\bibitem{Higginbotham2018}
Higginbotham A~P, Burns P~S, Urmey M~D, Peterson R~W, Kampel N~S, Brubaker B~M,
  Smith G, Lehnert K~W and Regal C~A 2018 {\em Nat. Phys.\/} {\bf 14}
  1038--1042 ISSN 17452481

\bibitem{Witmer2017a}
Witmer J~D, Valery J~A, Arrangoiz-Arriola P, Sarabalis C~J, Hill J~T and
  Safavi-Naeini A~H 2017 {\em Sci. Rep.\/} {\bf 7} 1--7 ISSN 20452322

\bibitem{Leuthold2009}
Leuthold J, Freude W, Brosi J~m, Baets R, Dumon P, Biaggio I, Scimeca M~L,
  Diederich F, Frank B and Koos C 2009 {\em Proc. IEEE\/} {\bf 97} 1304--1316
  ISSN 0018-9219

\bibitem{Leuthold2013}
Leuthold J, Koos C, Freude W, Alloatti L, Palmer R, Korn D, Pfeifle J,
  Lauermann M, Dinu R, Wehrli S, Jazbinsek M, G{\"{u}}nter P, Waldow M,
  Wahlbrink T, Bolten J, Kurz H, Fournier M, Fedeli J~M, Yu H and Bogaerts W
  2013 {\em IEEE J. Sel. Top. Quantum Electron.\/} {\bf 19} 114 -- 126 ISSN
  1077260X

\bibitem{Koos2016}
Koos C, Leuthold J, Freude W, Kohl M, Dalton L, Bogaerts W, Giesecke A~L,
  Lauermann M, Melikyan A, Koeber S, Wolf S, Weimann C, Muehlbrandt S, Koehnle
  K, Pfeifle J, Hartmann W, Kutuvantavida Y, Ummethala S, Palmer R, Korn D,
  Alloatti L, Schindler P~C, Elder D~L, Wahlbrink T and Bolten J 2016 {\em J.
  Light. Technol.\/} {\bf 34} 256--268 ISSN 07338724

\bibitem{Dalton2010}
Dalton L~R, Sullivan P~A and Bale D~H 2010 {\em Chem. Rev.\/} {\bf 110} 25--55
  ISSN 00092665

\bibitem{Kieninger2018}
Kieninger C, Kutuvantavida Y, Elder D~L, Wolf S, Zwickel H, Blaicher M, Kemal
  J~N, Lauermann M, Randel S, Freude W, Dalton L~R and Koos C 2018 {\em
  Optica\/} {\bf 5} 739 ISSN 2334-2536

\bibitem{Weis1985}
Weis R and Gaylord T 1985 {\em Appl. Phys. A Mater. Sci. Process.\/} {\bf 37}
  191--203 ISSN 07217250

\bibitem{Korn2013}
Korn D, Palmer R, Yu H, Schindler P~C, Alloatti L, Baier M, Schmogrow R,
  Bogaerts W, Selvaraja S, Lepage G, Pantouvaki M, Wouters J, Verheyen P, {Van
  Campenhout} J, Absil P, Baets R, Dinu R, Koos C, Freude W and Leuthold J 2013
  {\em Opt. Express\/} {\bf 21} 13219--13227 ISSN 1094-4087

\bibitem{Alloatti2014}
Alloatti L, Palmer R, Diebold S, Pahl K~P, Chen B, Dinu R, Fournier M, Fedeli
  J~M, Zwick T, Freude W, Koos C and Leuthold J 2014 {\em Light Sci. Appl.\/}
  {\bf 3} 5--8 ISSN 20477538

\bibitem{Wolf2018}
Wolf S, Zwickel H, Hartmann W, Lauermann M, Kutuvantavida Y, Kieninger C,
  Altenhain L, Schmid R, Luo J, Jen A~K, Randel S, Freude W and Koos C 2018
  {\em Sci. Rep.\/} {\bf 8} 1--13 ISSN 20452322

\bibitem{Miura2017}
Miura H, Qiu F, Spring A~M, Kashino T, Kikuchi T, Ozawa M, Nawata H, Odoi K and
  Yokoyama S 2017 {\em Opt. Express\/} {\bf 25} 28643 ISSN 1094-4087

\bibitem{Park2016}
Park D, Yun V, Luo J, Jen A~Y and Herman W 2016 {\em Electron. Lett.\/} {\bf
  52} 1703--1705 ISSN 00036951

\bibitem{Tsang2010}
Tsang M 2010 {\em Phys. Rev. A\/} {\bf 81} 063837 ISSN 1050-2947

\bibitem{Tsang2011}
Tsang M 2011 {\em Phys. Rev. A\/} {\bf 84} 1--8 ISSN 10502947

\bibitem{Aspelmeyer2014}
Aspelmeyer M, Kippenberg T~J and Marquardt F 2014 {\em Rev. Mod. Phys.\/} {\bf
  86} 1391--1452 ISSN 15390756

\bibitem{Safavi-Naeini2011a}
Safavi-Naeini A~H and Painter O 2011 {\em New J. Phys.\/} {\bf 13} 013017 ISSN
  1367-2630

\bibitem{Wang2012}
Wang Y~D and Clerk A~A 2012 {\em New J. Phys.\/} {\bf 14} ISSN 13672630

\bibitem{Joannopoulos2008}
Joannopoulos J~D, Johnson S~G, Winn J~N and Meade R~D 2008 {\em {Photonic
  Crystals}\/} 2nd ed (Princeton University Press)

\bibitem{Huang2012}
Huang S, Luo J, Jin Z, Zhou X~H, Shi Z and Jen A~K~Y 2012 {\em J. Mater.
  Chem.\/} {\bf 22} 20353--20357 ISSN 09599428

\bibitem{Heni2017}
Heni W, Haffner C, Elder D~L, Tillack A~F, Fedoryshyn Y, Cottier R, Salamin Y,
  Hoessbacher C, Koch U, Cheng B, Robinson B, Dalton L~R and Leuthold J 2017
  {\em Opt. Express\/} {\bf 25} 2627 ISSN 1094-4087

\bibitem{Chen2008a}
Chen H, Chen B, Huang D, Jin D, Luo J~D, Jen A~K and Dinu R 2008 {\em Appl.
  Phys. Lett.\/} {\bf 93} 1--4 ISSN 00036951

\bibitem{Zhang2015}
Zhang X, Hosseini A, Subbaraman H, Wang S, Zhan Q, Luo J, Jen A~K, Chung C~j,
  Yan H, Pan Z, Nelson R~L, Lee C~Y and Chen R~T 2015 {\em Terahertz, RF,
  Millimeter, Submillimeter-Wave Technol. Appl. VIII\/} {\bf 9362} 93620O ISSN
  1996756X

\bibitem{Degallaix2013}
Degallaix J, Flaminio R, Forest D, Granata M, Michel C, Pinard L, Bertrand T
  and Cagnoli G 2013 {\em Opt. Lett.\/} {\bf 38} 2047 ISSN 0146-9592

\bibitem{Witmer2016b}
Witmer J~D, Hill J~T and Safavi-Naeini A~H 2016 {\em Opt. Express\/} {\bf 24}
  5876 ISSN 1094-4087

\bibitem{Deotare2009}
Deotare P~B, McCutcheon M~W, Frank I~W, Khan M and Lon{\v{c}}ar M 2009 {\em
  Appl. Phys. Lett.\/} {\bf 94} 12--15 ISSN 00036951

\bibitem{Goban2015}
Goban A, Hung C~L, Hood J~D, Yu S~P, Muniz J~A, Painter O and Kimble H~J 2015
  {\em Phys. Rev. Lett.\/} {\bf 115} 1--5 ISSN 10797114

\bibitem{Benedikovic2014}
Benedikovic D, Cheben P, Schmid J~H, Xu D~X, Lapointe J, Wang S, Halir R,
  Ortega-Mo{\~{n}}ux A, Janz S and Dado M 2014 {\em Laser Photonics Rev.\/}
  {\bf 8} L93--L97 ISSN 18638899

\bibitem{McKenna2019}
McKenna T~P, Patel R~N, Witmer J~D, {Van Laer} R, Valery J~A and Safavi-Naeini
  A~H 2019 {\em Opt. Express\/} {\bf 27} 28782 ISSN 10944087

\bibitem{Wu1999}
Wu Y, Gamble H~S, Armstrong B~M, Fusco V~F and {Carson Stewart} J~A 1999 {\em
  IEEE Microw. Guid. Wave Lett.\/} {\bf 9} 10--12 ISSN 10518207

\bibitem{Mattis1958}
Mattis D~C and Bardeen J 1958 {\em Phys. Rev.\/} {\bf 111} 412--417 ISSN
  0031899X

\bibitem{Visser}
de~Visser P 2014 {\em {Quasiparticle dynamics in aluminium superconducting
  microwave resonators}\/} Ph.D. thesis TU Delft

\bibitem{Dieterle2016}
Dieterle P~B, Kalaee M, Fink J~M and Painter O 2016 {\em Phys. Rev. Appl.\/}
  {\bf 6} 014013 ISSN 2331-7019

\bibitem{Soltani2017}
Soltani M, Zhang M, Ryan C, Ribeill G~J, Wang C and Loncar M 2017 {\em Phys.
  Rev. A\/} {\bf 96} 1--9 ISSN 24699934

\bibitem{Rochman2019}
Rochman J, Bartholomew J~G, Craiciu I, Wang C, Xie T, Kindem J~M, Schwab K and
  Faraon A 2019 {\em Opt. InfoBase Conf. Pap.\/} {\bf Part F128-} 7--8

\bibitem{Owen1972}
Owen C~S and Scalapino D~J 1972 {\em Phys. Rev. Lett.\/} {\bf 28} 1559--1561
  ISSN 00319007

\bibitem{Zmuidzinas2012}
Zmuidzinas J 2012 {\em Annu. Rev. Condens. Matter Phys.\/} {\bf 3} 169--214
  ISSN 1947-5454

\bibitem{Rothwarf1967}
Rothwarf A and Taylor B~N 1967 {\em Phys. Rev. Lett.\/} {\bf 19} 27--30 ISSN
  00319007

\bibitem{Kaplan1976}
Kaplan S, Chi C~C, Langenberg D~N, Chang J~J, Jafarey S and Scalapino D~J 1976
  {\em Phys. Rev. B\/} {\bf 14} 4854--4873 ISSN 01631829

\bibitem{Barends2011}
Barends R, Wenner J, Lenander M, Chen Y, Bialczak R~C, Kelly J, Lucero E,
  O'Malley P, Mariantoni M, Sank D, Wang H, White T~C, Yin Y, Zhao J, Cleland
  A~N, Martinis J~M and Baselmans J~J~A 2011 {\em Appl. Phys. Lett.\/} {\bf 99}
  99--101 ISSN 00036951

\bibitem{Lang2003}
Lang K~M, Nam S, Aumentado J, Urbina C and Martinis J~M 2003 {\em IEEE Trans.
  Appl. Supercond.\/} {\bf 13} 989--993 ISSN 10518223

\bibitem{Samkharadze2016}
Samkharadze N, Bruno A, Scarlino P, Zheng G, Divincenzo D~P, Dicarlo L and
  Vandersypen L~M 2016 {\em Phys. Rev. Appl.\/} {\bf 5} 1--7 ISSN 23317019

\bibitem{Annunziata2010a}
Annunziata A~J, Santavicca D~F, Frunzio L, Catelani G, Rooks M~J, Frydman A and
  Prober D~E 2010 {\em Nanotechnology\/} {\bf 21} ISSN 09574484

\bibitem{Hazard2019}
Hazard T~M, Gyenis A, {Di Paolo} A, Asfaw A~T, Lyon S~A, Blais A and Houck A~A
  2019 {\em Phys. Rev. Lett.\/} {\bf 122} ISSN 10797114

\bibitem{Shearrow2018}
Shearrow A, Koolstra G, Whiteley S~J, Earnest N, Barry P~S, Heremans F~J,
  Awschalom D~D, Shirokoff E and Schuster D~I 2018 {\em Appl. Phys. Lett.\/}
  {\bf 113} ISSN 00036951

\bibitem{Yemane2017}
Yemane Y~T, Sowa M~J, Zhang J, Ju L, Deguns E~W, Strandwitz N~C, Prinz F~B and
  Provine J 2017 {\em Supercond. Sci. Technol.\/} {\bf 30} ISSN 13616668

\bibitem{Masluk2012}
Masluk N~A, Pop I~M, Kamal A, Minev Z~K and Devoret M~H 2012 {\em Phys. Rev.
  Lett.\/} {\bf 109} 1--5 ISSN 00319007

\bibitem{Bell2012}
Bell M~T, Sadovskyy I~A, Ioffe L~B, Kitaev A~Y and Gershenson M~E 2012 {\em
  Phys. Rev. Lett.\/} {\bf 109} 1--10 ISSN 00319007

\bibitem{Fink2016}
Fink J~M, Kalaee M, Pitanti A, Norte R, Heinzle L, Davan{\c{c}}o M, Srinivasan
  K and Painter O 2016 {\em Nat. Commun.\/} {\bf 7} 1--10 ISSN 20411723

\bibitem{Mohan1999}
Mohan S~S, Hershenson M~D~M, Boyd S~P and Lee T~H 1999 {\em IEEE J. Solid-State
  Circuits\/} {\bf 34} 1419--1420 ISSN 00189200

\bibitem{Norte2015}
Norte R~A 2015 {\em {Nanofabrication for On-Chip Optical Levitation,
  Atom-Trapping, and Superconducting Quantum Circuits}\/} Ph.D. thesis
  California Institute of Technology

\end{thebibliography}

\providecommand{\newblock}{}

\end{document}